\documentclass{article}

\usepackage{arxiv}
\usepackage[utf8]{inputenc}
\usepackage[T1]{fontenc}
\usepackage{hyperref}
\usepackage{url}
\usepackage{booktabs}
\usepackage{amsfonts}
\usepackage{nicefrac}
\usepackage{microtype}
\usepackage{graphicx}
\usepackage{subfig}
\usepackage{float}
\usepackage{geometry}
\usepackage{color, colortbl}
\usepackage{rotating}
\usepackage{subfig}
\usepackage{tikz}
\usepackage[all]{xy}
\usetikzlibrary{positioning} %Needed for the basic examples
\usetikzlibrary{arrows,decorations.pathmorphing,shapes} %Needed for the crazy example
\usepackage{tikz-cd}
\usepackage{xcolor}
\usepackage{animate}
\usetikzlibrary{shapes.geometric, arrows}
\usepackage{amsmath}
\usepackage[export]{adjustbox}
\usepackage[ruled,vlined]{algorithm2e}

\newcommand{\vect}[1]{\boldsymbol{#1}}
\newcommand{\blaveros}[1]{\textcolor[rgb]{0,0,0}{#1}}
\newcommand{\blue}[1]{\textcolor[rgb]{0,0,0}{#1}}
\newcommand{\blau}[1]{\textcolor[rgb]{0,0,1}{#1}}
\newcommand{\red}[1]{\textcolor[rgb]{1,0,0}{#1}}

\usepackage[backend=biber,style=numeric-comp,sorting=none]{biblatex}
\title{\blue{Functional Connectivity via Total Correlation: \\ Analytical results in Visual Areas}}

\author{
    Qiang Li$^{1,2}$\\
    $^{1}$Image Processing Laboratory\\
    University of Valencia, Spain\\
    $^{2}$Tri-Institutional Center for Translational Research in Neuroimaging and Data Science (TReNDS)\\
    Georgia State University, Georgia Institute of Technology, and Emory University \\
    Atlanta, GA 30303, United States\\
  \texttt{qli27@gsu.edu} \\
 \And
    Greg Ver Steeg$^{3}$\\ 
    $^{3}$Computer Science \& Engineering\\
    University of California Riverside\\
    Riverside, CA 92507, United States\\
    \texttt{greg.versteeg@ucr.edu}\\
 \And
    Jesus Malo$^{1}$\\ 
    $^{1}$Image Processing Laboratory\\
    University of Valencia\\
    Valencia, 46980, Spain  \\
  \texttt{jesus.malo@uv.es} \\
}

\begin{document}

\definecolor {processblue}{cmyk}{0.96,0,0,0}
\maketitle

\vspace{-1cm}
\begin{abstract}

Recent studies invoke the superiority of the multivariate \emph{Total Correlation} concept over the conventional pairwise measures of functional connectivity in biological networks.
%That seminal work was restricted to show that empirical measures of \emph{Total Correlation} lead to connectivity patterns that differ from what is obtained using \emph{linear correlation} and \emph{Mutual Information}. 
Those seminal works certainly show that empirical measures of \emph{Total Correlation} lead to connectivity patterns that differ from what is obtained using the most popular measure, \emph{linear correlation},  or its higher order and nonlinear alternative \emph{Mutual Information}.
%However, beyond the obvious multivariate versus bivariate definitions, no theoretical insight on the benefits of Total Correlation was given.
However, they do not provide analytical results that explain the differences beyond the obvious multivariate versus bivariate definitions. Moreover, 
the accuracy of the empirical estimators could not be addressed directly because no controlled scenario with known analytical result was provided either. This point is critical because empirical estimation of information theory measures is always challenging.

As opposed to previous empirical approaches, in this work we present analytical results to prove the advantages of \emph{Total Correlation} over \emph{Mutual Information} to describe the functional connectivity. In particular, we do it in neural networks for early vision (retina-LGN-cortex) which are realistic but simple enough to get analytical results.
%Our neural models include three layers (retina, LGN, and V1 cortex) and one can control the connectivity among the nodes, within the cortex, and the eventual top-down feedback.
%In this setting (more than two multidimensional nodes),
%we derive analytical results for the multi-way \emph{Total Correlation} and for all possible pairwise \emph{Mutual Information} measures.
%These analytical results show that pairwise \emph{Mutual Information} cannot capture the effect of different intra-cortical inhibitory connections while the three-way \emph{Total Correlation} can.
The presented analytical setting is also useful to check empirical estimates of \emph{Total Correlation}. 
Therefore, once certain estimate can be trusted, one can explore the behavior with natural signals where the analytical results (that assume Gaussian signals), may not be valid.
In this regard, as applications (a) we explore the effect of connectivity and feedback in the analytical retina-LGN-cortex network with natural images, and (b) we assess the functional connectivity in visual areas V1-V2-V3-V4 from actual fMRI recordings.

\end{abstract}


\section{Introduction}

%% Functional connectivity was described by pair-wise measures
Functional connectivity in \blaveros{brain networks} goes beyond structural links: it is related to the way information is shared among \emph{multiple} neural  nodes\protect~\cite{Friston11,Lizier11}.
Quantifying the communication among \emph{multiple} neural regions is key to understand brain function.
However, \blue{the most popular measure of functional connectivity is the (linear and \emph{pairwise}) \emph{Pearson Correlation}\protect~\cite{Mohanty20}.
Of course, more general concepts such as \emph{Mutual Information}\protect~\cite{FeiFei09} have been proposed to capture the nonlinear relations between pairs of nodes, but still they cannot cope with more than two nodes simultaneously. For instance, \emph{Transfer Entropy}\protect~\cite{Schreiber00}, which is a variant of \emph{Directed Information}\protect~\cite{Gastpar13,Massey90} and reduces to classical linear \emph{Granger Causality} when dealing with auto-regressive signals\protect~\cite{Barnett09}, is also based on \emph{conditional Mutual Information} (conditioning on the past of the signals in \emph{two} nodes). Therefore all these measures also belong to the pairwise family too. The problem is that measures which are limited to pairwise comparisons should be applied many times to describe complex networks, and 
they may miss interactions among multiple nodes\protect~\cite{Battiston20}.}

%% Recent studies proposed the multi-variate Total Correlation, but these works were empirical
Recent studies proposed the use of Total Correlation as a way to overcome the intrinsic pairwise limitation of the conventional measures of functional connectivity in neuroscience\protect~\cite{Li22,QiangEntr22}.
\blue{Other recent works\protect~\cite{Herzog22,Gatica21} reason in the same \emph{multi-node} direction using variations of Total Correlation}.
The multivariate nature of Total Correlation,~$T$\protect~\cite{Watanabe60}
is a \emph{by-definition} advantage over Mutual Information,~$I$\protect~\cite{Cover06}. However, the aforementioned seminal works 
had a fundamental limitation: beyond the obvious multivariate definition of $T$,
no extra theoretical insight on its benefits over $I$ was given.
As a consequence of the lack of analytical models and results, the accuracy of the empirical estimators could not be addressed because no controlled scenario was considered either. This is an important limitation because the empirical estimation of information theoretic quantities is very challenging particularly in high dimensions. Note that dimensionality is crucial in complex networks where one may have to consider the response of nodes with many neurons responding over long periods of time. The situation is even worse when multiple nodes have to be considered.
Analytical results are crucial to trust and understand the differences in connectivity found using different information measures: the new measures are showing something new or the new trends simply come from biased estimations?. 

%% In contrast, here (our main contribution) is the analytical approach
The goal of this work is addressing the limitation of the empirical approaches in~\cite{Li22,QiangEntr22}:
we present \emph{analytical} results on the superiority of $T$ over $I$ in a specific context: the early visual brain. This focus on closed-form expressions restricts the range of comparisons but makes the conclusions solid. We do it through the consideration of simple but realistic analytical models of the retina-cortex pathway.

%% The retina-cortex model
\blaveros{The three-node model considered here (retina-LGN-V1) consists of the conventional linear receptive fields plus the biological version of batch-normalization: the  Divisive Normalization nonlinearity~\cite{Carandini94,Rust06,Carandini12,Martinez18,Martinez19}, and we 
consider variations with top-down feedback~\cite{Kandel91}.
There are several reasons to choose this kind of neural network: 
(1)~it reproduces the psychophysics of subjective image quality, as explicitly checked here following~\cite{Watson02,Malo10,Laparra10,Gomez19}, 
(2)~every layer has noisy neurons so that one can compute the part of the information that is lost along the neural pathway~\cite{Malo20,Malo22}, 
(3)~the interest of this class of models goes beyond visual neuroscience given the similarity of its linear+nonlinear structure with other sensory modalities~\cite{Schwartz01,AudioAdapt,Tashi23}, with popular networks in computer vision such as AlexNet~\cite{alexnet} or VGG~\cite{vgg16}, and with image coding algorithms~\cite{Watson94,Ahumada97,malo06a,camps-valls2008a}, and finally, 
(4)~the divisive normalization is a canonical computation in the brain~\cite{Carandini12} so it is important to develop descriptors that can capture its inhibitory connectivity.}

\begin{figure}[t!]
    \vspace{-0cm}
    \centering
\includegraphics[width=0.35\textwidth]{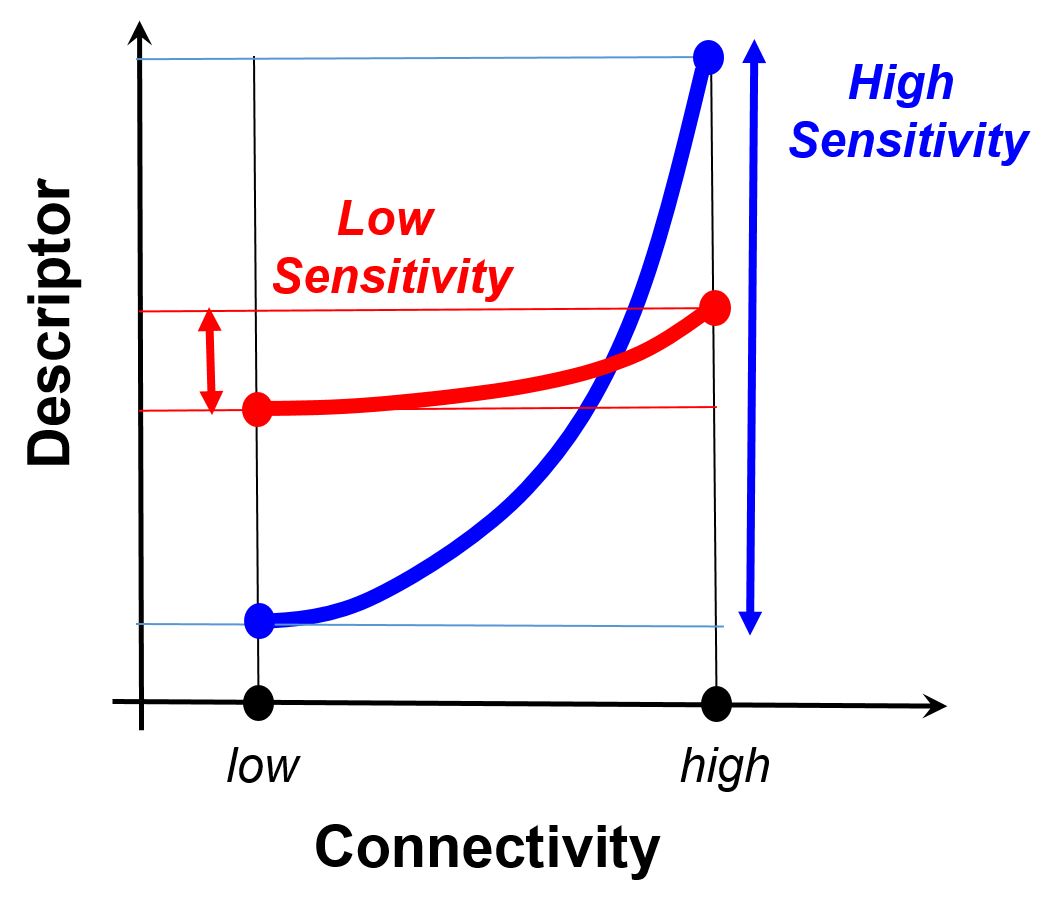}
\caption{\small{\textbf{The basic concept:}
    The best descriptor of functional connectivity is the one with bigger sensitivity (variation with connectivity). 
    The analytical results derived in this work show that \emph{Total Correlation} has bigger sensitivity to connectivity than \emph{Mutual Information}.}}
    \label{fig:goal}
\end{figure}

%% Descriptor comparison using the sensitivity
\blaveros{Descriptors of connectivity can be rated according to their \emph{sensitivity}. 
Note that a descriptor of a magnitude can be seen as an instrument to measure this magnitude. The sensitivity of an instrument is given by the slope of 
its response curve~\cite{Ivanov69}. In a linear instrument the sensitivity is the ratio between the output (response, or measure in the y-axis) and the input (stimulus in the x-axis). In regular instruments of measurement, both input and output may be subject to noise. In that case, high sensitivity implies highly noticeable response variations, which is convenient in presence of noise in the y-axis, but also implies amplification of the input, which is a problem in presence of noise in the x-axis. 
In contrast to this instrument metaphor, in the case of the descriptor of a magnitude, there is no noise in the x-axis (the magnitude of interest, e.g. connectivity, has certain value), and all the noise is in the y-axis (error in the estimation of the descriptor from the available data, e.g. the estimation of  information from the samples at different nodes). When noise (or error) is restricted to the y-axis, the sensitivity determines the minimum variation of the magnitude that can be noticed by the descriptor (on top of its inherent noise). In the same way, the larger the sensitivity of a descriptor, the more robust it is to the noise in the estimation (in the y-axis). In fact, if two descriptors have the same error, the one preferred is the one with bigger sensitivity~\cite{Mandel54}. This concept is illustrated in Fig.~\ref{fig:goal}, and the variation of the descriptor over a region of connectivity values (sensitivity) will be used to decide between descriptors.}

%% Main contributions
\blaveros{The contributions of this study are the following:}
\begin{itemize}

\item \blaveros{We derive expressions for the descriptors $T$ and $I$ depending on the feedforward and feedback structural connectivity of the retina-cortex pathway 
and on the properties of signal and noise.}

\item \blaveros{Our analytical results show that while $I$ is insensitive to some of the connectivity parameters, $T$ is always more sensitive to the connectivity in the retina-cortex pathway.
As opposed to previous empirical approaches, these analytical results explicitly show the superiority of $T$ over $I$ as a description of the connectivity in biologically plausible neural networks.}

\item \blaveros{The analytical results constitute a test-bed to check the accuracy of different empirical estimators for $T$ (or $I$). In this way, available estimators (as for instance~\cite{Laparra11,steeg2014NIPS,steeg17,steeg2015corex_theory,Marin-Franch13,szabo14}) can be reliably applied to real data where theoretical results are not available (for instance because the Gaussian assumption is no longer valid~\cite{Field96,Olshausen01,Malo06b,Malo10}).}

\item \blaveros{After checking the information theoretic measures in controlled scenarios, we use a recent fMRI dataset~\cite{Cichy21} to measure the information flow among deeper visual cortical areas and conjecture about their interactions/synergies. Furthermore, we suggested that \textit{data processing inequality} holds in the human vision cortex.}
    
\end{itemize}

\iffalse

In this work we derive expressions for the descriptors $T$ and $I$ depending on the feedforward and feedback structural connectivity and the properties of signal and noise. 
%As illustrated in Fig.~\ref{fig:goal}, the key is the \emph{sensitivity of the descriptor}: the bigger the variation of the descriptor in the range of explored connectivity the better. 
%With bigger sensitivity, the representation of the connectivity will be more robust to errors in the estimation of the descriptor. 
Our analytical results show that while $I$ is insensitive to some of the connectivity parameters, $T$ is always more sensitive to the connectivity in the retina-cortex pathway.
These analytical results explicitly show the superiority of $T$ over $I$ as a description of the connectivity in biologically plausible neural networks.

On the other hand, the presented analytical results constitute a test-bed to check the accuracy of different empirical estimators for $T$ (or $I$). In this way, available estimators (as for instance~\cite{Laparra11,steeg2014NIPS,steeg17,steeg2015corex_theory,Marin-Franch13,szabo14}) can be reliably applied to real data where theoretical results are not available (for instance because the Gaussian assumption is no longer valid~\cite{Field96,Olshausen01,Malo06b,Malo10}). 
Finally, in this paper we discuss the shared information between cortical areas using a recent fMRI dataset~\cite{Cichy21}.

\fi

The structure of the paper is as follows.
Section~\ref{SectionMaterials} (\emph{materials and methods}) describes the class of neural models considered throughout the work, and reviews the definitions of $I$ and $T$. Section~\ref{Anal_results} describes the theoretical results: we derive the expressions that describe the functional connectivity (using $I$ and $T$) in terms of the parameters of the networks. These analytical results consider both feedforward nonlinear networks, and networks with feedback.
Section~\ref{Anal_emp_results} shows a range of  experimental illustrations of the theory: it presents results of $T$ and $I$ computed with empirical estimators that can be compared to the theoretical predictions. 
Moreover, results for real signals (natural images and actual fMRI responses) are also presented here.
Finally, Section~\ref{Discussion} summarizes the results and discusses the implications of the work.
Appendices present supplementary material that can be omitted from the main text: \blaveros{Appendices~A and~B introduce the parameters of the specific vision models and their biological plausibility.} 
Appendix~C reviews the relations of $I$ and $T$ with the (more limited, but still widely used) classical linear correlation, and
%Appendix D gives the details of the fMRI data, and 
Appendix~D illustrates the variability of the empirical estimations of $I$ and $T$ with real signals.
\blaveros{Appendix~E empirically illustrates the insensitivity of other alternative measures of connectivity~\cite{Zhou09}.}

\section{Materials and Methods}
\label{SectionMaterials}
In this section we first introduce the notation of the standard analytical models of early vision that will be used throughout the work (based on a large body of evidences~\cite{DeAngelis97,Watson92,Hancock92,Carandini12,Esteve20,Brainard19,Watson97,Martinez18,Martinez19,Carandini94,Rust06}, and explicitly checked here using standard methods~\cite{Watson02,Laparra10,Malo10,Gomez19,Malo20}). Then, we review the definitions of the considered descriptors of connectivity ($T$, $I$, and \emph{linear correlation}) which are based on classical information theoretic concepts~\cite{Cover06,Watanabe60,Cardoso03}.

%On the one hand, the standard early vision networks formulated below are based on a large body of physiological and psychophysical facts~\cite{DeAngelis97,Watson92,Hancock92,Carandini12,Esteve20,Brainard19,Watson97,Martinez18,Martinez19}. 
%Moreover, we explicitly check the biological plausibility of the specific parameters used in this work following a standard method extensively used before~\cite{Watson02,Laparra10,Malo10,Gomez19,Malo20}.  
%On the other hand, the considered descriptors of connectivity are based on classical information theoretic concepts~\cite{Cover06,Watanabe60,Cardoso03}.

\subsection{Models of the retina-cortex pathway}
\label{Models}

Expanding and making explicit the multi-node scenario first considered in~\cite{Li22}, all the theoretical results of this work will be derived for the following \emph{early vision} setting that may include feedforward and feedback connections, as seen in this \blaveros{graphic diagram:}
\vspace{0.3cm}
\begin{equation}
\xymatrix{\mathrm{Retina} \ar[r] &  \mathrm{LGN} \ar[r]  & \ar@/^1.5pc/[ll] \ar@/_1.5pc/[l] \mathrm{V1}}
\end{equation}
\vspace{0.0cm}

In this \blaveros{diagram} the arrows represent structural connections between regions (or layers). Right-arrows represent feedforward flow of the visual information, and the left-arrows represent eventual feedback.

More specifically, the signal at the \emph{retina} will be represented by the $n$-dimensional random vector, $\mathbf{x}$, the signal at the \emph{LGN}, will be represented by the $n$-dimensional random vector, $\mathbf{y}$,
and the signal at the cortex will be represented by two $n$-dimensional random vectors, $\mathbf{e}$ and $\mathbf{z}$. 
In this way, the intra-cortical connectivity is represented by the communication between $\textbf{e}$ and $\textbf{z}$.
In the following \blaveros{diagram} the strength of the structural connections between layers $i$ and $j$ is represented by the variables, $c_{ij}$: 
\begin{equation}
\xymatrix{\mathbf{x} \ar[r]^{c_{xy}} &  \mathbf{y} \ar[r]^{c_{ye}}  & \mathbf{e} \ar[r]^{c_{ez}}  & \ar@/^1.5pc/[lll]^{c_{zx}} \ar@/_1.5pc/[ll]_{c_{zy}} \mathbf{z}}
\label{Framework}
\end{equation}

In the above setting, the study of functional connectivity through information-theoretic measures (such as $I$ or $T$) could be useful to describe the \emph{unknown} strengths, $c_{ij}$, from recordings of the neural signal done at the different nodes or layers.   
In this context, proper measures of statistical relation should be sensitive to $c_{ij}$. And, as illustrated in Fig.~\ref{fig:goal}, the bigger the sensitivity to the strength of the connections, the better.

\subsubsection{\emph{Model I}: Nonlinear and noisy model with focus on intra-cortical interactions}
\label{Models_no_recurrence}

Our first specific example of the retina-cortex framework outlined in \blaveros{diagram}~\ref{Framework}, which we will refer to as \emph{Model I}, tries to be analytically simple yet biologically plausible. To do so, this models includes: 
(a)~center-surround receptive fields in the LGN~\cite{DeAngelis97},
(b)~local-frequency receptive fields in the (linear) V1-cortex, approximated here as block-DCT functions~\cite{Watson92,Hancock92}, 
(c)~Divisive Normalization to model cortical nonlinearities~\cite{Carandini12}, and 
(d)~noise in each of the neural layers is scaled in a way compatible with the psychophysical results in~\cite{Esteve20} and the physiological model in~\cite{Brainard19}.

The class of networks under \emph{Model I} follows these equations:
\begin{eqnarray}
    \mathbf{x}(t) &=& \mathbf{s}(t) + \mathbf{n_x}(t) + \frac{c_{zx}}{c_{xy}\,c_{ye}} \, F^{-1} \cdot \mathbf{z}(t-\Delta t)  \nonumber \\[0.0cm]
    \mathbf{y}(t) &=& c_{xy} \, K \cdot \mathbf{x}(t) + \mathbf{n_y}(t) \,\,\,\,\,\,\,\,=\,\,\,\,\,\,\,\, c_{xy} \, F^{-1} \cdot \lambda_{\textrm{CSF}} \cdot F \cdot \mathbf{x}(t) + \mathbf{n_y}(t)  \label{Norecur} \\[0.2cm]
    \mathbf{e}(t) &=& c_{ye} \, F \cdot \mathbf{y}(t) + \mathbf{n_e}(t) \nonumber\\
    \mathbf{z}(t) &=& f\left( \mathbf{e}(t) \right) \,\,\,\,\,\,\,\,=\,\,\,\,\,\,\,\, \operatorname{sign}(\mathbf{e}(t)) \cdot \kappa \cdot \frac{|\mathbf{e}(t)|^\gamma}{b \,\, + \,\, c_{ez} \, H \cdot |\mathbf{e}(t)|^\gamma} \nonumber
\end{eqnarray}
where, the input to the system is the retinal image: the source vector $\mathbf{s} \in \mathbb{R}^n$, and its dimension $n$ corresponds to the number of photoreceptors. In the models considered in this work, the networks preserve the dimension of the signal\footnote{Preservation of dimension along the pathway is convenient but it doesnt reduce the generality neither biologically, the spatial subsampling affects the extrafovea, but not the fovea~\cite{Kandel91}, nor mathematically because changes of dimension could be addressed by the Jacobians of rectangular transforms~\cite{Li19}.}. 

The retinal signal, the vector $\mathbf{x} \in \mathbb{R}^n$, is influenced by the input image $\mathbf{s}$, but it is also affected by the white noise $\mathbf{n_x}$ and in this formulation, by a top-down feedback given by the term weighted by $c_{zx}$, that describes the strength of this feedback connection.  
Due to the eventual variations in the input and the feedback, all the multivariate signals may depend on time, $t$.
We will come back to the feedback term once we introduce the frequency meaning of vector $\mathbf{z}$.

The signal at the LGN is described by the vector $\mathbf{y} \in \mathbb{R}^n$. The matrix $K$ contains the center-surround receptive fields of LGN~\cite{DeAngelis97}. According to the relation between these receptive fields and the Contrast Sensitivity Function (CSF)~\cite{Uriegas94,Atick92,Li22csfs}, we implement them using a local-frequency transform (basis in the matrix $F$), a diagonal matrix with CSF-related weights, $\lambda_{CSF}$, and coming back to the spatial domain using $F^{-1}$.
The LGN signal is also affected by white noise through $\mathbf{n_y}$.

The (intermediate) linear signal at the V1-cortex, $\mathbf{e}$, is computed from the LGN signal through a set of local-frequency receptive fields in the matrix $F$. This linear signal is also affected by the white noise $\mathbf{n_e}$.

The (final) nonlinear signal at V1, $\mathbf{z}$, results from a Divisive Normalization transform, $f(\cdot)$, of the outputs of the linear receptive fields at the previous intermediate layer, $\mathbf{e}$. Note that the division, the exponent, and the absolute values in $f(\cdot)$ are Hadamard (element-wise) operations~\cite{Martinez18}, and the  matrix $H$ in the denominator represents the interaction between the neurons of the previous cortical layer $\mathbf{e}$.
%Specifically, the elements $H_{kl}$ represent the interaction between the $k$-th and the $l$-th neurons. 
Specifically, the intra-cortical connectivity between the $k$-th and the $l$-th neurons is represented by $c_{ez} H_{kl}$. In this way, the $k$-th row of $H$, $H_{kl} \,\,\, \forall l=1,\ldots,n$, describes how the responses of the neighbor linear neurons, $e_l$, affect the nonlinear response of the $k$-th neuron, $z_k$.
This interaction is assumed to be local in space and frequency~\cite{Watson97,Martinez18,Martinez19}.
And $c_{ez}$ controls the global strength of all these local interactions. 

Finally, a comment on the top-down feedback terms in the first equation. The Divisive Normalization changes the relative magnitude of the responses $z_i$ but the rough qualitative meaning of the responses in $\mathbf{z}$ is still given by the (local-frequency) receptive fields in $F$. Therefore, the $F^{-1}$ matrix in the top-down feedback term in the first equation of the system just converts the previous cortical response $\mathbf{z}(t-\Delta t)$ back into the spatial domain (where the input images $\mathbf{s}$ are).
Additionally, the top-down term has been scaled by the other connectivity strengths ($c_{xy}$ and $c_{ye}$) just to keep the scale of the feedback term comparable to the source independently of the (arbitrary) gains introduced along the retina-cortex path. In this way the effective weight of the feedback term only depends on $c_{zx}$. 

%In Section~\ref{Models_plausibility} we will see that the above elements 
In \emph{Appendix A} we show the specific values chosen for the receptive fields, the frequency selectivity, and the patterns of intra-cortical connectivity. We also illustrate the responses arising in these networks when stimulated by natural images.

In \emph{Appendix B} we show that the above elements (the considered layers and noise levels) are biologically realistic.
In particular, this architecture
explains human opinion in visual distortion psychophysics. In this regard, the intra-cortical connectivity in the Divisive Normalization transform is particularly critical. Therefore, eventual measures of the statistical relation between neural nodes should be sensitive to this intra-cortical connectivity.

The parameters that control the feedforward structural connections between retina, LGN, and the linear V1, (i.e. the strengths $c_{xy}$ and $c_{ye}$) actually control the size of the signal with regard to the noise, and hence their functional role is quite evident: the bigger the signal compared to the noise, the stronger the information flow from one node/layer to the next. However, the role of the intra-cortical interaction $c_{ez} H$ is more interesting. There is a large body of literature that suggests that the role of the denominator in Divisive Normalization is capturing-and-removing the statistical relations between the responses of the linear local-frequency sensors~\cite{Wainwright00,Schwartz01,malo06a,Malo06b,Malo10,Coen12}. 

The first set of analytical results derived in Section~\ref{res_anal_1} shows that $T$ is sensitive to this intra-cortical connectivity, while the sensitivity of $I$ to these intra-cortical connections is equal to zero. 
These connections have major biological relevance~\cite{Carandini12,Hyvarinen01,Ma08}
but are also important in artificial networks~\cite{Hepburn2020,Hernandez23}.
This is an analytical example of the genuine superiority of the Total Correlation over the conventional Mutual Information. 

\subsubsection{\emph{Model II}: Linear noisy model with focus on feedback}
\label{Models_recurrence}

\emph{Model II} is just a variation of \emph{Model I} intended to simplify the analytical study of feedback. 
The convenience of this variation will become apparent in Section~\ref{Anal_results} when we derive the analytical results.
By comparing the Eqs.~\ref{Norecur}~of \emph{Model I} and Eqs.~\ref{Recur} of \emph{Model II} it is easy to see that our second class of networks is just a linear version of the first where we disregarded the Divisive Normalization. Specifically, in the last equation of \emph{Model~II} the cortical nonlinearity $f(\cdot)$ has been substituted by a trivial identity, $\mathbb{I}$, and the input cortical signal is scaled by the strength $c_{ez}$ with regard to the inner noise $\mathbf{n_z}$, which was not present before:
\begin{eqnarray}
    \mathbf{x}(t) &=& \mathbf{s}(t) + \mathbf{n_x}(t) + \frac{c_{zx}}{c_{xy}\,c_{ye}\,c_{ez}} \, F^{-1} \cdot \mathbf{z}(t-\Delta t)  \nonumber \\[0.0cm]
    \mathbf{y}(t) &=& c_{xy} \, K \cdot \mathbf{x}(t) + \mathbf{n_y}(t) \,\,\,\,\,\,\,\,=\,\,\,\,\,\,\,\, c_{xy} \, F^{-1} \cdot \lambda_{\textrm{CSF}} \cdot F \cdot \mathbf{x}(t) + \mathbf{n_y}(t) \label{Recur} \\[0.2cm]
     \mathbf{e}(t) &=& c_{ye} \, F \cdot \mathbf{y}(t) + \mathbf{n_e}(t) \nonumber\\[0.2cm]
    \mathbf{z}(t) &=& c_{ez} \, \mathbb{I} \cdot \mathbf{e}(t) + \mathbf{n_z}(t) \nonumber
\end{eqnarray}
\vspace{0.0cm}

In the setting described by \emph{Model II} the information about the input image (or source $\mathbf{s}$) flows through the feedforward links while being contaminated by the noise injected at each layer.
However, for the slow-varying inputs described above, part of the source is injected back into the retinal signal. 
As a result, the scenario in \emph{Model II} is convenient to analyze the joint effect of the strength of the feedforward links and the feedback links. For example, one may study the effect of the intra-cortical connectivity $c_{ez}$ (that scales the signal wrt the inner noise) together with the strength of the feedback $c_{zx}$ that reinforces the presence of the source at the retina. From a naive perspective, increasing $c_{ez}$ and $c_{zx}$ seems to lead to an increase of the Signal-to-Noise ratio in all the responses. Analytical results of information-theoretic descriptors can confirm or refute this intuition and provide a tool to understand a variety of situations. 

The second set of analytical results derived in  Section~\ref{res_anal_2} show that while 
$T$ strongly depends on the feedforward and feedback strengths
$c_{ez}$ and $c_{zx}$, the sensitivity of $I$ is smaller. 
In this case, the sensitivity of $I$ is just smaller (not zero) but 
the substantial difference in sensitivities (in a biologically plausible recurrent scenario) illustrates the conceptual superiority of $T$ over the conventional $I$.

%%%%%%%%%%%%%%%%%
\subsection{Background on Mutual Information and Total Correlation}
\label{background}

\blue{Here we recall the definitions of the descriptors compared in this work (\emph{Mutual Information}~\cite{Cover06} and 
\emph{Total Correlation}~\cite{Watanabe60}), in terms of \emph{Entropy}:}
\begin{eqnarray}
 T(\mathbf{x},\mathbf{y},\mathbf{z}) & = & \left( \sum_{i=1}^n h(x_i) + h(y_i) + h(z_i) \right)- h(\mathbf{x},\mathbf{y},\mathbf{z}) \label{defT}\\
      I(\mathbf{x},\mathbf{y}) & = & h(\mathbf{x}) + h(\mathbf{y}) - h(\mathbf{x},\mathbf{y}) \label{defI}
\end{eqnarray}
where $h(\cdot)$ stands for the (univariate or joint) entropy of the corresponding (scalar or vector) variables. The relation of these variables with the (more limited but still widely used) 2nd-order linear correlation is detailed in \emph{Appendix C}.
The biggest conceptual advantage of $T$ over $I$ and \emph{linear correlation} is that it can handle relations among more than two nodes at the same time. \blaveros{Note that the definition in Eq.~\ref{defT} is trivially extended in presence of an arbitrary number of nodes.}
Moreover, similarly to $I$, $T$ can capture nonlinear relations as opposed to 2nd order linear correlation. However, $T$ is different from $I$. Note that even in the case of just two nodes, $T(\mathbf{x},\mathbf{y})\neq I(\mathbf{x},\mathbf{y})$ because, for multivariate nodes, $T$ considers the redundancy among the coefficients (or neurons) of each node, which is disregarded by $I$. This difference is key when the signals in each layer are not independent, which is the more interesting situation in visual neuroscience and also in computer vision.    

As joint and marginal entropy are easily computed for Gaussian signals from the covariance matrices or from the marginal variances~\cite{Cover06}, Eqs.~\ref{defT} and~\ref{defI} imply that, if variables are Gaussian, analytical results are straightforward.
This is the case in \emph{Model II}, but, due to the nonlinearity, it is not the case in \emph{Model I}.

%%%%%%%%%%%%%%%%%

\section{Analytical results: $T$ and $I$ in terms of intra-layer connectivity and feedback}\label{Anal_results}

Here we present results for \emph{Model I} and \emph{Model II} which address different interesting situations that may happen in natural or artificial neural nets: (i)~nonlinear intra-layer connectivity, and (ii)~feedback or recurrence. In order to simplify the analytical tractability, in each case we focus on a specific feature of the models, either the nonlinearity (in \emph{Model I}) or the feedback-recurrence (in \emph{Model II}).

For simplicity in the notation, in this Section we omit the temporal variation of the signals. Nevertheless, as discussed in Section~\ref{Discussion} in the paragraph \emph{Temporal delays can be incorporated in the theory}, that is not a major problem because the properties of $T$ used in the proofs do not depend on time. 

For both models (\emph{I} and \emph{II}) analytical tractability is simple if one considers Gaussian signals. The Gaussian assumption for natural images has been acknowledged as a too rough approximation both in Visual Neuroscience~\cite{Field96,Olshausen01,Malo06b,Malo10} and in Image Processing~\cite{Portilla01,Gutierrez06}. However, in this section we are going to take this assumption for the sake of analytical tractability. In the experimental section we will compare the results with (synthetic) Gaussian signals and natural inputs. The Gaussian assumption is appropriate and illustrative in this case because (as shown below using a trustable empirical estimator) results for natural images are (1) similar to the Gaussian results, and more important for this work, (2) they confirm the superiority of the description using $T$ also for natural signals. 

\subsection{$T$ and $I$ as descriptors of intra-cortical connectivity (\emph{Model I})}
\label{res_anal_1}

As stated above, 
for simplicity, 
%in this section, on top of the Gaussian assumption mentioned above, we will also consider $c_{zx}=0$ in our nonlinear 
we consider a Gaussian input, $\vect{s}$, and a version of \emph{Model I} focused on the nonlinearity. This means $c_{zx}=0$, so we leave feedback for the results of \emph{Model II} in Section \ref{res_anal_2}.

With these assumptions, the variables $\mathbf{x}$,  $\mathbf{y}$, and $\mathbf{e}$ are Gaussian because they are sum of linearly-transformed Gaussian variables plus white Gaussian noises. However, the Divisive Normalization nonlinearity $f(\cdot)$ implies that the variable $\mathbf{z}$ is non-Gaussian. In this setting, expressions for $T$ and $I$ involving $\mathbf{z}$ (where the intra-cortical connectivity is) require the application of specific properties of these magnitudes under transforms of the random variables.

\textbf{\emph{The Total Correlation does depend on intra-cortical connectivity:}} In order to get an analytical result for $T(\mathbf{x},\mathbf{y},\mathbf{z})$, lets concatenate the variables that represent the considered nodes into column vectors of dimension $3n$: $\mathbf{a} = [\mathbf{x};\mathbf{y};\mathbf{e}]$, and $\mathbf{a'} = [\mathbf{x};\mathbf{y};\mathbf{z}] = [\mathbf{x};\mathbf{y};f(\mathbf{e})]$, and consider,
\begin{equation}
    \mathbf{a} \xrightarrow{\,\,\,\,\mathcal{F}\,\,\,\,}\mathbf{a'}
    \nonumber
\end{equation}
where we are interested in computing $T(\mathbf{a'})$.
In this situation, one may use the following property of the variation of Total Correlation when the variables undergo a transformation $\mathcal{F}$~\cite{Lyu09,Malo20}:
\begin{equation}
    \Delta T(\mathbf{a},\mathbf{a'}) = T(\mathbf{a})-T(\mathbf{a'}) = \sum_i^{3n} h(a_i) -  \sum_i^{3n} h(a'_i) + \frac{1}{2} \mathbb{E}_\mathbf{a} \bigl\{ \, log|\nabla_\mathbf{a} \mathcal{F}^\top \cdot \nabla_\mathbf{a} \mathcal{F}| \, \bigr\}
    \label{variaT}
\end{equation}
where $\mathbb{E}_\mathbf{a} \bigl\{ \cdot \bigr\}$ is the average over the samples $\mathbf{a}$. 
Then, taking into account that, 
\begin{equation}
   \nabla_\mathbf{a} \mathcal{F} = \left(
   \begin{array}{cc}
    \mathbb{I} \,\,&\,\,  0 \\[0.2cm]
    0  \,\,&\,\,  \nabla_{\mathbf{e}} f \\
   \end{array}\right)
   \nonumber
\end{equation}
and considering that $T(\vect{a}) =T(\vect{x},\vect{y},\vect{e})$ only depends on Gaussian variables and hence with known entropy in terms of the covariance matrix\footnote{If $\mathbf{x}$ is a Gaussian variable, its entropy in \emph{nats} is $h(\mathbf{x}) = \frac{1}{2}log |2\pi e \Sigma^x|$ where $\Sigma^x$ is the covariance of $\mathbf{x}$~\cite{Cover06}.}, we obtain the desired result (in \emph{nats}):
\begin{equation}
    T(\vect{x},\vect{y},\vect{z}) = \frac{1}{2} \!\sum_i^{3n} log (\Sigma^{a}_{ii})  - \frac{1}{2} log |\Sigma^{a}| - \frac{n}{2} - \frac{n}{2}log(2 \pi) - \frac{1}{2} log |\Sigma^e| 
                    + \sum_{i=1}^n h(z_{i})
                   - \frac{1}{2} \mathbb{E}_{\vect{e}}\{ \, log|\nabla_{\vect{e}} f \cdot \nabla_{\vect{e}} f^\top| \, \}
                   \label{TheorT1}
\end{equation}
where the covariance matrices $\Sigma^{e}$ and $\Sigma^{a}$ do not depend on the intra-cortical connectivity, because they only depend on $\vect{x}$,  $\vect{y}$, and $\vect{e}$:
\begin{equation}
    \Sigma^{a}=\Sigma^{xye}=\left(\begin{array}{ccc}
    \Sigma^{x} &  c_{xy} \cdot \Sigma^{x} \cdot K^{\top} & c_{ye} \cdot c_{xy} \cdot \Sigma^{x} \cdot (F \cdot K)^{\top} \\
    c_{xy} \cdot K \cdot \Sigma^{x} & \Sigma^{y} & c_{ye} \cdot \Sigma^{y} \cdot F^{\top} \\
    c_{ye} \cdot c_{xy} \cdot F \cdot K  \cdot \Sigma^{x} & c_{ye} \cdot F \cdot \Sigma^{y} & \Sigma^{e}
    \end{array}\right)
    \nonumber
\end{equation}
but, according to~\cite{Martinez18}, $\nabla_{\vect{e}} f$ does depend on the intra-cortical connectivity due to the interactions in the Divisive Normalization, $c_{ez}$ and $H$:
\begin{equation}
\nabla_{\vect{e}} f = \mathbb{D}_{sign(\vect{e})} \cdot \mathbb{D}^{-1}_{\big( b + c_{ez} \cdot H\cdot|\vect{e}| \big)}\cdot[\mathbb{I} - c_{ez} \cdot \mathbb{D}_{\vect{z}} \cdot H] \cdot \mathbb{D}_{ \big( \gamma \,\, sign(\vect{e}) |\vect{e}|^{\gamma-1} \big)} 
%\cdot \mathbb{D}_{sign(\vect{e})} 
\label{Eq.deltaf}
\end{equation}
where $\mathbb{D}_{\vect{v}}$ is a diagonal matrix with the vector $\vect{v}$ in the diagonal.

Eqs.~\ref{TheorT1} and \ref{Eq.deltaf} explicitly show that $T(\vect{x},\vect{y},\vect{z})$ \emph{does} depend on the intra-cortical connectivity.

Another way to see the dependence with the intra-cortical connectivity consist of identifying these two terms in Eq.~\ref{TheorT1}: the (Gaussian) $T(\vect{x},\vect{y},\vect{e})$, using the definition in Eq.~\ref{defT}, and the variation of $T$ under the transform $\mathbf{z}=f(\mathbf{e})$, using the property in Eq.~\ref{variaT}. By doing that, it is easy to see that: \begin{equation}
       T(\vect{x},\vect{y},\vect{z}) = \Big( T(\vect{x},\vect{y},\vect{e}) - T(\vect{e}) \Big) + T(\vect{z})  
       \label{TheorT1b}
\end{equation}
where the term in the parenthesis obviously does not depend on the intra-cortical connectivity (because $\vect{x}$, $\vect{y}$ and $\vect{e}$ are previous to that interaction), but $T(\mathbf{z})$ \emph{does} depend on the Divisive Normalization.

\textbf{\emph{The Mutual Information does not capture the effect of intra-cortical connectivity:}} This is easy to see using the following property: the mutual information is invariant to non-singular differentiable transforms of the random vectors~\cite{Kraskov04}:
\begin{equation}
    I(\mathbf{a},f(\mathbf{b})) = I(\mathbf{a},\mathbf{b})
    \label{invarianceI}
\end{equation}
This property is easy to see by considering that $I(\mathbf{a},\mathbf{b})$ measures the KL-divergence between the densities $p(\mathbf{a},\mathbf{b})$ and $p(\mathbf{a})p(\mathbf{b})$~\cite{Cover06}. Taking 
into account that the Jacobian that appears in the variation of the probability under transforms~\cite{Stark94} is compensated (in the integral of the KL-divergence) by the change of the differential volume, one gets the invariance.

As a result, no \emph{pairwise} measure $I$ involving $\vect{x}$, $\vect{y}$, and $\vect{z}$ depends on the intra-cortical connectivity: 
\begin{equation}
    \begin{array}{l}
    I(\vect{x}, \vect{y})=\frac{1}{2} \log \left|\Sigma^{x}\right|+\frac{1}{2} \log \left|\Sigma^{y}\right|-\frac{1}{2} \log \left|\Sigma^{xy}\right| \\ \\
    I(\vect{x},\vect{z})=I(\vect{x},f(\vect{e}))=I(\vect{x},\vect{e})=\frac{1}{2} \log \left|\Sigma^{x}\right|+\frac{1}{2} \log \left|\Sigma^{e}\right|-\frac{1}{2} \log \left|\Sigma^{xe}\right| \\ \\
    I(\vect{y},\vect{z})=I(\vect{y},f(\vect{e}))=I(\vect{y},\vect{e})=\frac{1}{2} \log \left|\Sigma^{y}\right|+\frac{1}{2} \log \left|\Sigma^{e}\right|-\frac{1}{2} \log \left|\Sigma^{ye}\right| \end{array}
    \label{gaussmi}
\end{equation}
where,
\begin{equation}
    \Sigma^{xy}=\left(\begin{array}{cc}
    \Sigma^{x} &  c_{xy} \cdot \Sigma^{x} \cdot K^{\top}\\
    c_{xy} \cdot K \cdot \Sigma^{x} & \Sigma^{y}
    \end{array}\right)
    \nonumber
\end{equation}

\begin{equation}
    \Sigma^{xe}=\left(\begin{array}{cc}
    \Sigma^{x} &  c_{ye} \cdot c_{xy} \cdot \Sigma^{x} \cdot (F \cdot K)^{\top}\\
    c_{ye} \cdot c_{xy} \cdot F \cdot K \cdot \Sigma^{x}  & \Sigma^{e}
    \end{array}\right)
    \nonumber
\end{equation}

\begin{equation}
    \Sigma^{ye}=\left(\begin{array}{cc}
    \Sigma^{y} & c_{ye} \cdot \Sigma^{y} \cdot F^{\top} \\
    c_{ye} \cdot F \cdot \Sigma^{y}  & \Sigma^{e}
    \end{array}\right)
    \nonumber
\end{equation}

Therefore, we proved an important advantage of $T$: in the biologically plausible \emph{Model I}, 
%(Pearson $\rho = 0.84$ with human opinion), 
Eq.~\ref{gaussmi} means that the conventional $I$ measures \emph{do not capture} the intra-cortical connectivity, 
which is critical to explain psychophysics (see Appendix B).
On the contrary, Eqs.~\ref{TheorT1} and \ref{Eq.deltaf} explicitly show that $T$ \emph{does} depend on the intra-cortical connectivity.

\subsection{$T$ and $I$ as descriptors of feedback (\emph{Model II})}
\label{res_anal_2}

In \emph{Model II} there is no nonlinearity so, if the source $\mathbf{s}$ is Gaussian and so are the noises injected at the different layers, all the variables (in the forward pass) will be Gaussian including $\mathbf{z}$. Then, the considered feedback from $\mathbf{z}$ to $\mathbf{x}$ just injects an extra Gaussian variable back into $\vect{x}$. As a result, $\mathbf{x}$ will be Gaussian too for any strength of the feedback.
For slow-varying inputs (as natural images at the retina) the feedback signal (coming from the past) is not totally independent of the current value of the source, so the covariance at the retina is not the sum of the covariance matrices of the separate terms in the sum in the first equation of \emph{Model II}. However, this does not modify the Gaussian assumption.

All these considerations imply that the definitions in terms of entropy given in Eqs.~\ref{defT} and~\ref{defI} can be applied together with the expression of the entropy for Gaussian signals that only depends on the corresponding covariance matrices. As a result, in order to make explicit the dependence on feedforward and feedback connectivity one only has to consider all possible covariance matrices, which is what we list below for \emph{Model II}.

Assuming that signal and noise are not correlated, the covariance matrices of the signal at each isolated layer are:
\begin{eqnarray}
    \Sigma^{x}&=&\mathbb{E} \bigl\{x \cdot x^\top \bigr\} = \Sigma^s + \Sigma^{n_x} + \Big( \frac{c_{zx}}{c_{xy}c_{ye}c_{ez}}\Big)^2 F^{-1} \cdot \Sigma^z \cdot F^{-1}{}^\top + \frac{c_{zx}}{c_{xy}c_{ye}c_{ez}} M(s,z) \nonumber \\
    \Sigma^{y}&=&c_{xy}^{2} \cdot K \cdot \Sigma^{x} \cdot K^\top + \sigma^2(n_y) \, \mathbb{I}  \label{one_II}\\
    \Sigma^{e}&=&c_{ye}^{2} \cdot F \cdot \Sigma^{y} \cdot F^{T} + \sigma^2(n_e) \, \mathbb{I}  \nonumber \\
    \Sigma^{z}&=&c_{ez}^{2} \cdot \Sigma^{e}  + n_{e}^{2} \cdot \mathbb{I}_{d} \nonumber
\end{eqnarray}
where $M(s,z)$ is a symmetric matrix that describes the relation between s and z (they are not independent), and it is given by: $M(s,z) = F^{-1} \cdot \mathbb{E}\bigl\{s \cdot z^\top \bigr\} + \big( F^{-1} \cdot \mathbb{E}\bigl\{s \cdot z^\top \bigr\} \big)^\top$.

Additionally, the covariance matrices of \emph{two} concatenated vectors that have not been given in Section~\ref{res_anal_1} are:
\begin{equation}
    \Sigma^{xz}=\left(\begin{array}{cc}
    \Sigma^{x} & c_{ye} \cdot c_{xy}\cdot c_{ez} \cdot \Sigma^{x} \cdot (F\cdot K)^{\top}  \\
    c_{ye} \cdot c_{xy}\cdot c_{ez} \cdot F \cdot K \cdot \Sigma^{x}   & \Sigma^{z}
    \end{array}\right)
    \nonumber
\end{equation}

\begin{equation}
    \Sigma^{yz}=\left(\begin{array}{cc}
    \Sigma^{y} & c_{ye} \cdot c_{ez}\cdot \Sigma^{y} \cdot F^{\top} \\
    c_{ye} \cdot c_{ez} \cdot F \cdot \Sigma^{y}   & \Sigma^{z}
    \end{array}\right)
    \label{two_II}
\end{equation}

\begin{equation}
    \Sigma^{ez}=\left(\begin{array}{cc}
    \Sigma^{e} & c_{ez} \cdot \Sigma^{e} \\
    c_{ez} \cdot \Sigma^{e}  & \Sigma^{z}
    \end{array}\right)
    \nonumber
\end{equation}

Similarly, the covariance matrices of \emph{three} and \emph{four} concatenated vectors that have not been given in Section~\ref{res_anal_1} are:

\begin{equation}
    \Sigma^{xyz}=\left(\begin{array}{ccc}
    \Sigma^{x} &  c_{xy} \cdot \Sigma^{x} \cdot K^{\top} & c_{ye} \cdot c_{xy} \cdot c_{ez} \cdot \Sigma^{x} \cdot (F \cdot K)^{\top} \\
    c_{xy} \cdot K \cdot \Sigma^{x} & \Sigma^{y} & c_{ye} \cdot c_{ez} \cdot \Sigma^{y} \cdot F^{\top} \\
    c_{ye} \cdot c_{xy} \cdot c_{ez} \cdot F \cdot K  \cdot \Sigma^{x} & c_{ye} \cdot c_{ez} \cdot F \cdot \Sigma^{y} & \Sigma^{z}
    \end{array}\right)
    \nonumber
\end{equation}

\begin{equation}
    \Sigma^{xez}=\left(\begin{array}{ccc}
    \Sigma^{x} &  c_{xy} \cdot c_{ye} \cdot \Sigma^{x} \cdot (F \cdot K)^{\top} & c_{ye} \cdot c_{xy} \cdot c_{ez} \cdot \Sigma^{x} \cdot (F \cdot K)^{\top} \\
    c_{xy} \cdot c_{ye} \cdot F \cdot K \cdot \Sigma^{x} & \Sigma^{e} & c_{ez} \cdot \Sigma^{e} \\
    c_{ye} \cdot c_{xy} \cdot c_{ez} \cdot F \cdot K  \cdot \Sigma^{x} & c_{ez} \cdot \Sigma^{e} & \Sigma^{z}
    \end{array}\right)
    \label{more_II}
\end{equation}

\begin{equation}
    \Sigma^{xyez}=\left(\begin{array}{cccc}
    \Sigma^{x} &  c_{xy} \cdot \Sigma^{x} \cdot K^{\top} & c_{ye} \cdot c_{xy} \cdot \Sigma^{x} \cdot (F \cdot K)^{\top} & c_{ye} \cdot c_{xy} \cdot c_{ez} \cdot \Sigma^{x} \cdot (F \cdot K)^{\top} \\
    
    c_{xy} \cdot K \cdot \Sigma^{x} & \Sigma^{y} & c_{ye} \cdot \Sigma^{y} \cdot F^{\top} & c_{ye} \cdot c_{ez} \cdot \Sigma^{y} \cdot F^{\top}\\
    
    c_{ye} \cdot c_{xy} \cdot F \cdot K  \cdot \Sigma^{x} & c_{ye} \cdot F \cdot \Sigma^{y} & \Sigma^{e} & c_{ez} \cdot \Sigma^{e} \\
    
    c_{ye} \cdot c_{xy} \cdot c_{ez} \cdot F \cdot K \cdot \Sigma^{x} & c_{ye} \cdot c_{ez} \cdot F \cdot \Sigma^{y} & c_{ez} \cdot \Sigma^{e} & \Sigma^{z} \\
    
    \end{array}\right)
    \nonumber
\end{equation}
Given the matrices in Eqs.~\ref{one_II}-\ref{more_II}, in  \emph{Model II} both variables $T$ and $I$ depend on the intra-cortical connectivity $c_{ez}$ and on the feedback $c_{zx}$. However, the sensitivity of the descriptors is not that obvious from these equations plugged into Eqs.~\ref{defT} and \ref{defI}. 
Therefore, in order to figure out which descriptor is better (which one is more sensitive) one should consider specific values of the parameters (e.g. what we consider in Appendices A and B), and compute $T$ and $I$ in a range of connectivity values.

We do that in the next experimental section where we find that, 
%similarly to what happens with intra-cortical connectivity in \emph{Model I}, 
in \emph{Model II}, our descriptor, $T$, is substantially more sensitive than $I$ to the feedback, $c_{zx}$, and the intra-cortical connectivity, $c_{ez}$. And this happens both for Gaussian signals and also for natural images.

\section{Empirical results}
\label{Anal_emp_results}

In this experimental section\footnote{Code and data at \blue{\texttt{http://isp.uv.es/docs/CODE\_connectivity.zip}},  \blue{\texttt{Samples.tar.gz}}, and \blue{\texttt{DATA\_connect\_2.zip}}} we address the following points:
\begin{itemize}

\item We use the theoretical expressions to illustrate the behaviors of $T$ and $I$, both in the case where the superiority of $T$ is analytically obvious (as in Eqs.~\ref{TheorT1}-\ref{TheorT1b} versus Eqs.~\ref{gaussmi} for the intra-cortical connectivity in \emph{Model I}), and in the case where the behavior is not easy to see directly from Eqs.~\ref{one_II}-\ref{more_II} plugged into Eqs.~\ref{defT}-\ref{defI} (in Model II). 
In these experiments we use Gaussian sources with the same mean and covariance as natural images and the model parameters discussed in Appendices A and B.

\item We confirm the theoretical results presented in Section~\ref{Anal_results} for both models (\emph{I} and \emph{II}) through a specific empirical estimator of $T$ and $I$~\cite{Laparra11,Laparra20} that has been already used in visual neuroscience~\cite{Gomez19,Malo20}. This empirical confirmation of the theory uses
sets of $0.5 \cdot 10^5$ Gaussian samples injected into the models (\emph{I} and \emph{II}), and then, the empirical estimator is applied to the responses of the models. 
% Gaussian sources with the same mean and covariance matrices as natural images. 
Incidentally, the presented pair \emph{theory-data} is a good test-bed for empirical estimators of $T$ and $I$.

\item We explore how the empirical estimations of $T$ and $I$ behave for natural (non-Gaussian) images where, in principle, the theory would not be applicable. We also use sets of $0.5 \cdot 10^5$ natural image patches and the same variations of \emph{Model I} and \emph{Model II}.

\item We explore the behavior of $T$ and $I$ in real fMRI signals from  cortical regions V1, V2, V3, V4 responding to natural images so that we can discuss possible connectivity schemes.

\end{itemize}

The structure of this section is as follows: 
(1) We describe the experimental issues: 
the empirical estimator, 
the natural and the synthetic image data, and the
computational issues associated with the theoretical expressions. 
(2) We present $T$ and $I$ surfaces for different intra-cortical connectivity 
$c_{ez}$ and $\alpha_H$ that controls $H$ in \emph{Model I}.
(3) We present $T$ and $I$ surfaces for different feedforward and feedback connectivity $c_{ez}$ and $c_{zx}$ in \emph{Model II}.
%Above we compare the theoretical results with the empirical results obtained both for Gaussian samples and natural samples.
Finally, (4) we present the empirical estimations of $T$ and $I$ from real fMRI recordings.

\subsection{Empirical estimator, image data, and computational issues}
\label{Methods}

\textbf{Empirical estimation of $T$ and $I$ from samples:} here we use the \emph{Rotation-Based Iterative Gaussianization} (RBIG). This method, originally proposed for PDF estimation~\cite{Laparra11}, is able to transform data following any multivariate PDF into data that follows a unit-covariance multivariate Gaussian. In this way, RBIG is useful to estimate the redundancy among coefficients because it accumulates the variations in redundancy while transforming the original dataset into the final Gaussian dataset where all coefficients are independent. The advantages of RBIG with regard to other  information estimators~\cite{szabo14,Marin-Franch13} has been shown in~\cite{Johnson19,Laparra20,Malo20}.
RBIG has also been used in visual neuroscience to check the \emph{Efficient Coding Hypothesis} in Wilson-Cowan networks~\cite{Gomez19}, in Divisive Normalization networks~\cite{Malo20}, and in color appearance networks~\cite{Malo2019FI}. However, any other empirical estimator of $T$ and $I$ from samples~\cite{szabo14,Marin-Franch13,steeg2014NIPS,steeg17,steeg2015corex_theory} could be used in the experiments below.

\begin{figure}[b!]
\vspace{-0.25cm}
    \centering
    \includegraphics[width=0.78\textwidth, height=10cm]{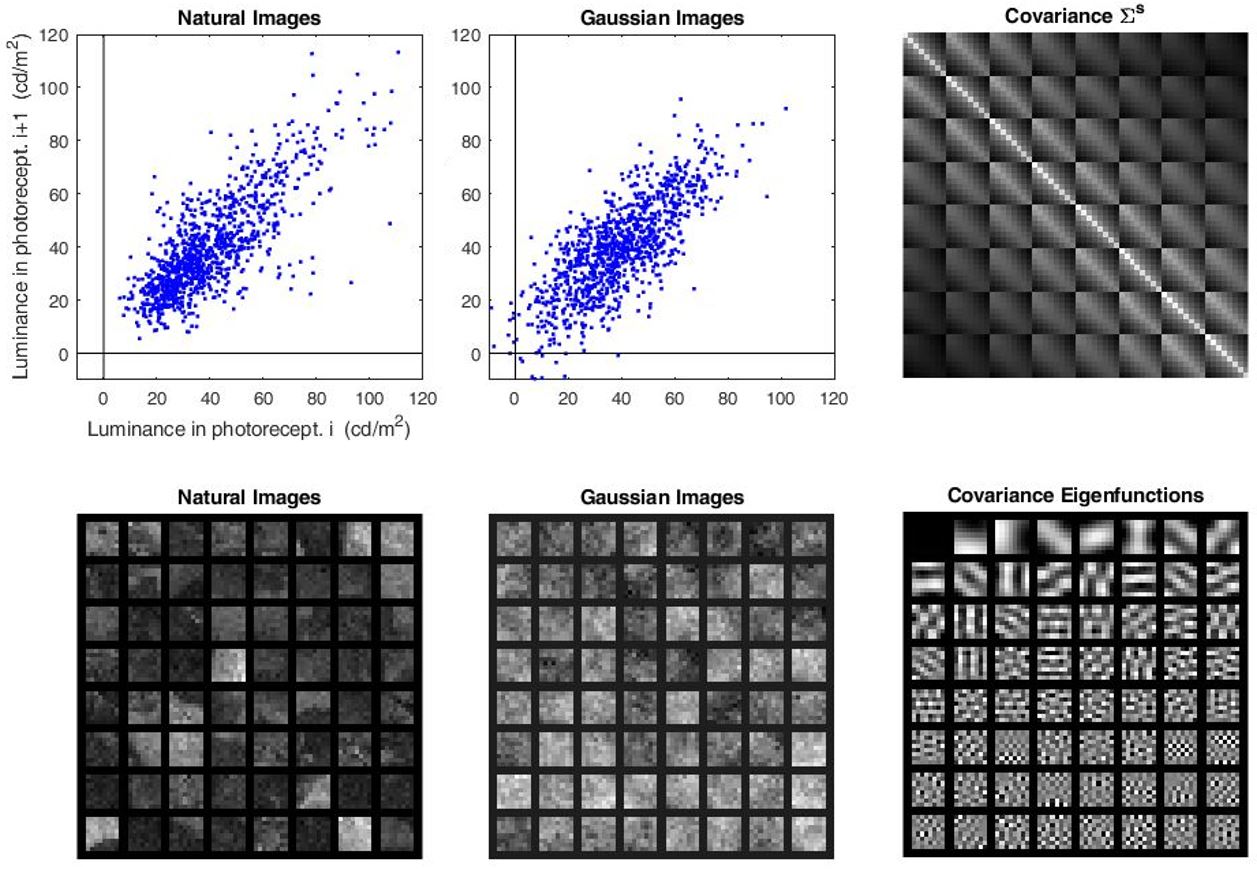}
    \caption{\small{\textbf{Natural and synthetic image data} (the source $\vect{s}$). The bottom-left mosaic shows illustrative samples from the colorimetrically-calibrated databases IPL and Barcelona. The top-left scatter plot illustrates the joint PDF of the luminance at neighbor photoreceptors. Images and scatter plot show the (non-Gaussian) bias towards low-luminance, and the spatial smoothness of the signal (predominance of low spatial frequency). The non-diagonal nature of the covariance matrix (at the top-right) captures the spatial smoothness, and its eigenfunctions (bottom-right) are similar to the frequency analyzers in the cortex models (in Fig.~\ref{fig:parameters}). 
    The order of the functions according the eigenvalue confirms the low-frequency nature of the signal. The central column shows Gaussian samples with the same mean and covariance.}}
    \label{fig:image_data}
\end{figure}

\textbf{Natural and synthetic image data:} In the experiments we used $0.5 \cdot 10^5$ image patches of size $8\times8$, i.e. $n=64$, randomly taken from the luminance component of two colorimetrically-calibrated datasets: the IPL dataset~\cite{Laparra12,Gutmann14}, and the Barcelona dataset~\cite{Parraga09}.
In the IPL dataset only images under the CIE D65 (daylight-like) illuminant were considered. The two datasets were linearly scaled so that the average luminance in both was equal to 40 $cd/m^2$. This separate global normalization ensures that image patches from both sets are equivalent and can be safely mixed. Then, we randomly extracted the samples $0.25\cdot10^5$ from each dataset, and we computed the covariance from this joint set of $0.5 \cdot 10^5$ samples: see $\Sigma^s$ in Fig.~\ref{fig:image_data}.
This matrix, $\Sigma^s$, is the starting point of all the theoretical results presented in Section~\ref{Anal_results}.
Our data has the classical covariance of the luminance in natural images (see for instance~\cite{Epifanio03}), which is diagonalized by DCT-like basis functions (see Fig.~\ref{fig:image_data}, consistently with~\cite{Clarke81,Hancock92,Gutmann14}).
Then, we generated $0.5 \cdot 10^5$ Gaussian vectors of dimension $n=64$ with the mean and covariance of the natural samples. Of course, both sets (natural and synthetic) are not the same (as can be seen in Fig.~\ref{fig:image_data}, consistently with~\cite{Field96,Olshausen01}).
Then, we inject the synthetic and natural samples through \emph{Model I} and \emph{Model~II} to get the corresponding responses $\vect{x}$, $\vect{y}$, $\vect{e}$, and $\vect{z}$, for the range of connectivity values considered in Section~\ref{Models}.

\textbf{Computational issues:} 
All the analytical results (e.g. Eq.~\ref{TheorT1}) depend on the computation of determinants of large matrices (either covariance matrices or the Jacobian $\nabla_{\vect{e}} f^\top \cdot \nabla_{\vect{e}} f$).
The computation of determinants in high-dimensional scenarios is very prone to divergences to $0$ or $\infty$. Therefore, it is better to avoid its computation: given the fact that the considered matrices, $A$, are symmetric (either $\Sigma$ or $\nabla_{\vect{e}} f^\top \cdot \nabla_{\vect{e}} f$), they are diagonalizable by an orthonormal transform (with unit determinant). Therefore, it holds $log |A| = \sum_{i=1}^d log(\lambda_i)$ where $\lambda_i$ are the eigenvalues of $A$ (whatever the dimension $d\times d$ of the matrix $A$). Note that this sum is more robust than the naive computation of the determinant.

\subsection{Results for $I$ and $T$ in terms of nonlinear intra-cortical connectivity (\emph{Model I})}
\label{Anal_emp_results_no_rec}

\begin{figure}[b!]
    \centering
    \includegraphics[width=0.8\textwidth, height=11cm]{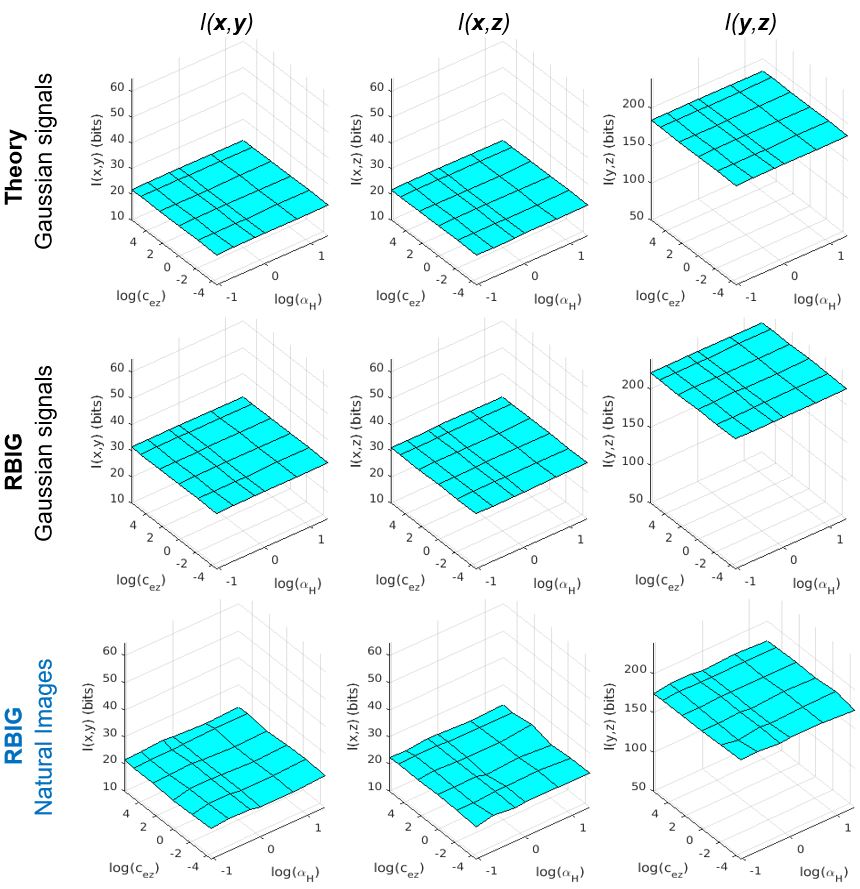}
    \caption{\small{\textbf{Mutual Information does not describe intra-cortical connectivity in \emph{Model~I}}. Plots of $I$ as a function of intra-cortical connectivity for Gaussian signals (theory and RBIG estimates), and empirical results for natural images.}}
    \label{fig:I_forward}
\end{figure}

Figure~\ref{fig:I_forward} shows the results of Mutual Information for different intra-cortical connectivity scenarios in the nonlinear \emph{Model~I}. Specifically, we show 
(a)~the theoretical results for Gaussian signals,
(b)~the empirical results computed with RBIG for Gaussian signals, and 
(c)~the empirical results computed with RBIG for natural signals. 

We see two basic trends in the results (both in the theory and in the empirical estimations):
\begin{enumerate}
    \item As predicted by the theory, Mutual Information is \emph{totally insensitive} to the differences in intra-cortical connectivity. Therefore, this pairwise measure is not a good descriptor of connectivity for this kind of nonlinearity, which is canonical in neural computation~\cite{Carandini12}.

    \item $I(\vect{x},\vect{y}) \approx I(\vect{x},\vect{z}) \ll I(\vect{y},\vect{z})$. This could be expected because the shared information is reduced with the noise introduced in each layer and $\sigma(\vect{n_y})\gg\sigma(\vect{n_e})$, and no noise is introduced in $\vect{z}$, i.e. $f(\cdot)$ is invertible. Therefore, more information is lost between $\vect{x}$ and inner layers (either $\vect{y}$ or $\vect{z}$), than the information lost between $\vect{y}$ and $\vect{z}$, which have an almost invertible relation: only a small fraction of bits is lost due to $\vect{n_e}$. 
    % The values are (T_I_feedforward.m  & T_I_feedforward_nature_images.m):
    %               Ixy    Ixz   Iyz
    %  Theor        21.8   21.7  183.6
    %  RBIG         31.3   31.2  221.4
    %  RBIG_nat     22.6   22.7  176.2

\end{enumerate}

It is important to note that these global trends in the theory are consistently confirmed by the empirical estimations.
Beyond a small bias (overestimation) in $I_{\mathrm{RBIG}}$, it identifies the substantially bigger connection between $\vect{y}$ and $\vect{z}$ rather than between $\vect{x}$ and inner layers.
Moreover, $I_{\mathrm{RBIG}}$ is also constant over the range of nonlinear connectivity values.

Interestingly, the empirical results for natural images also follow these trends even though the signals are no longer Gaussian. In this case, the non-Gaussianity only introduces a reduction in the  $I_{\mathrm{RBIG}}$ estimates and a small variation over the explored models, which is negligible in terms of describing changes in the connectivity.

% (which is quite usual in the visual pathway~\cite{Carandini12})

\begin{figure}[t!]
    \vspace{-1cm}
    \centering
    \includegraphics[width=0.95\textwidth, height=7cm]{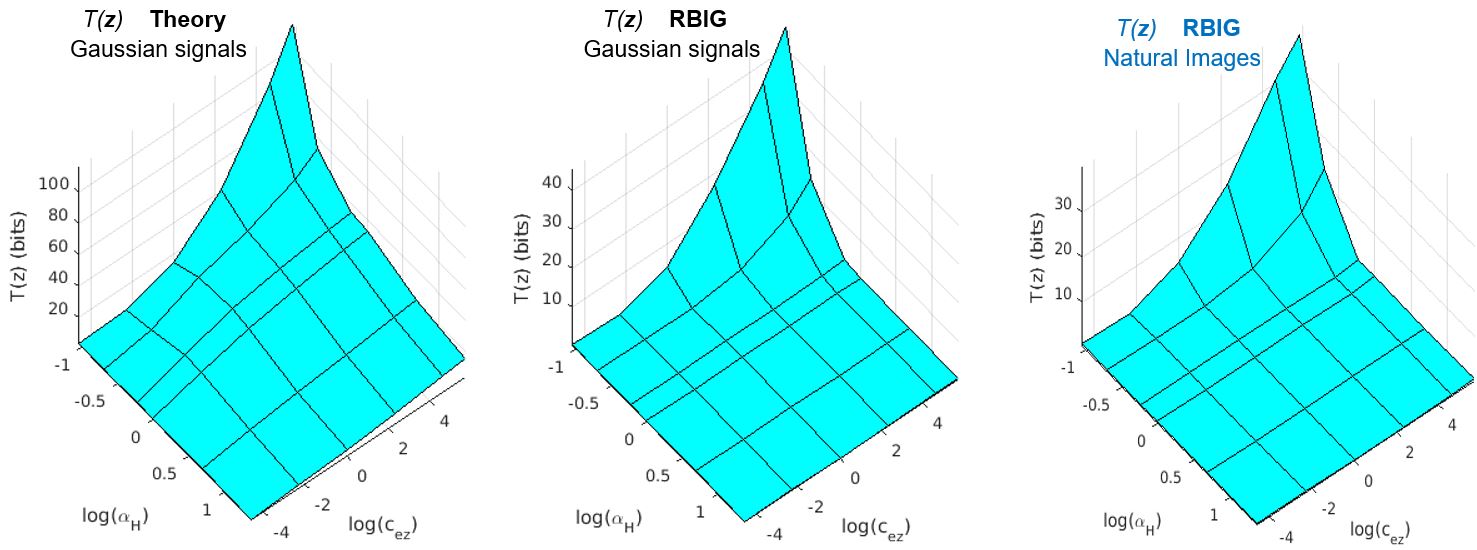}
    \caption{\small{\textbf{Total Correlation does capture variations in intra-cortical connectivity in \emph{Model~I}}.
    Plots of $T(\vect{z})$ as a function of intra-cortical connectivity for Gaussian signals (theory and RBIG estimates), and empirical results for natural images.}}
    \label{fig:T_forward}
\end{figure}

Figure~\ref{fig:T_forward} shows the part of $T(\vect{x},\vect{y},\vect{z})$ that depends on the nonlinear connectivity: $T(\vect{z})$ according to Eq.~\ref{TheorT1b}. 
In this case, as opposed to $I$, the \emph{Total correlation} strongly depends on the intra-cortical connectivity.

Again, (beyond a subestimation bias in RBIG) the general trend of the empirical estimations over the connectivity range confirms the theoretical predictions. The non-Gaussianity of natural signals does not introduce major deviations in the trend of the surface. 

A technical comment on the estimation of $T(\vect{z})$: as the variables  $\vect{z}=f(\vect{e})$ are non-Gaussian, and this non-Gaussianity is particularly strong in some regions of the explored domain of connectivity, it is important to use a large number of iterations in the Gaussianization algorithm to get a good estimate of $T$. In particular here we used 500 iterations.

\subsection{Results for $I$ and $T$ in terms of feedforward and feedback connectivity (\emph{Model II})}
\label{Anal_emp_results_rec}

As in the recurrent \emph{Model II} the interpretation of the analytical results is more complicated, here the values are given in a relative scale with regard to their maximum so that the sensitivity of the different descriptors can be fairly compared.
Moreover, the variation of each descriptor, $\Delta_I$ or $\Delta_T$, both in percentage and in bits, is also given. 
As the explored range of feedforward and feedback values is the same for each descriptor, $\Delta_I$  and $\Delta_T$ are good measures of the sensitivity to the considered variation of the connectivity.

\begin{figure}[!ht]
    \begin{centering}
    \hspace{-1.7cm}
    \includegraphics[width=1.175\textwidth,height=14cm]{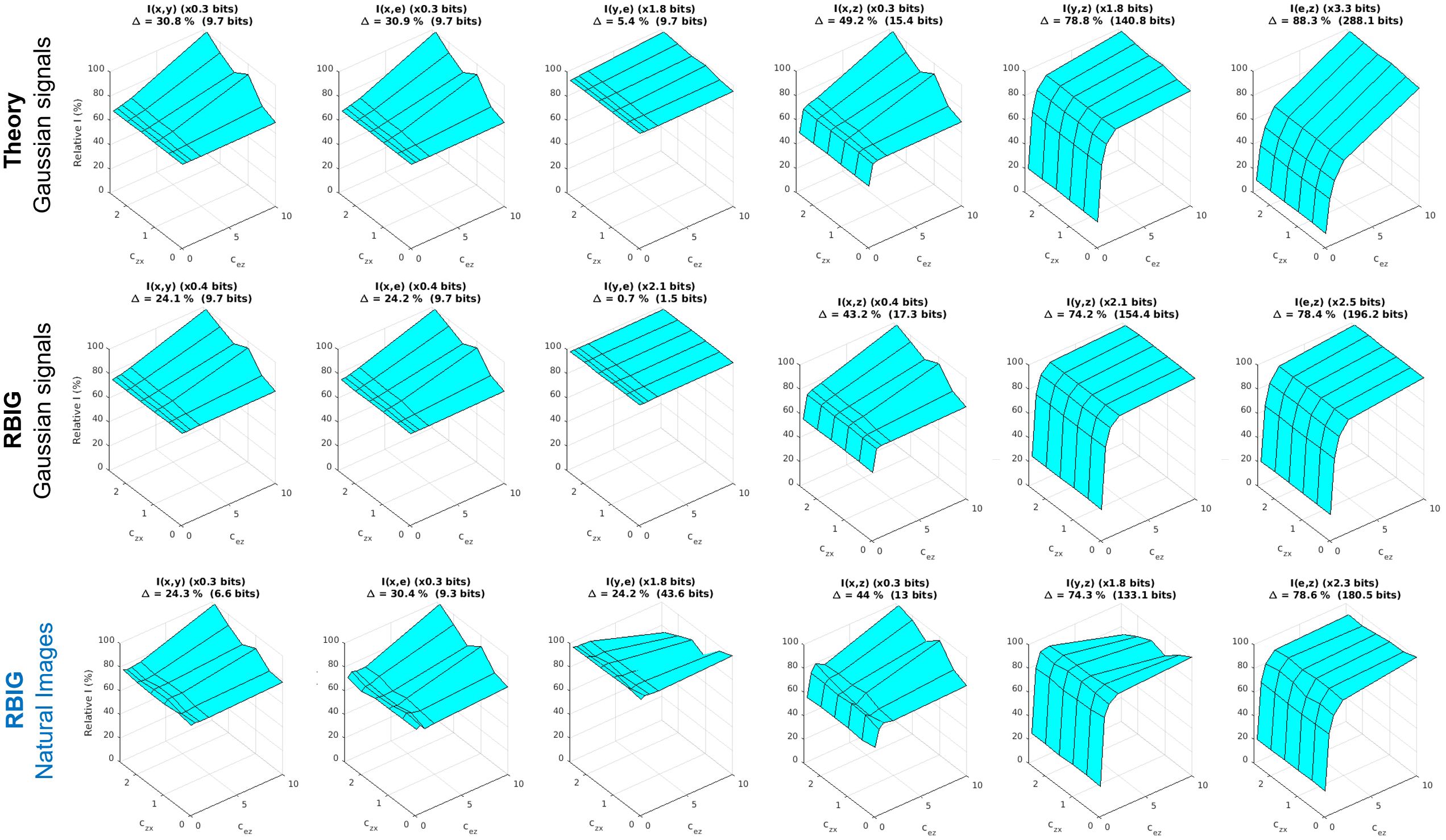}        
    \end{centering}
    \caption{\small{\textbf{Mutual Information has mild dependence with feedforward and feedback connectivity in \emph{Model~II}}. Plots of $I$ as a function of the feedforward connectivity, $c_{ez}$, and feedback, $c_{zx}$, for Gaussian signals (theory and RBIG estimates), and empirical results for natural images.
    The plots display relative values of $I$ in percentage with regard to the maximum together with a factor (e.g. $\times0.3$ in the top-left plot) that allows to express this percentage in absolute values (in bits). Moreover, the plots display the variation (in bits) of the considered descriptor over the range of connectivity values (e.g. $\Delta = 9.7$ bits in the top-left plot). This is a measure of the sensitivity of the descriptor.}}
    \label{fig:I_feedback}
\end{figure}

\begin{figure}[!ht]
    \centering
    \includegraphics[width=0.75\textwidth]{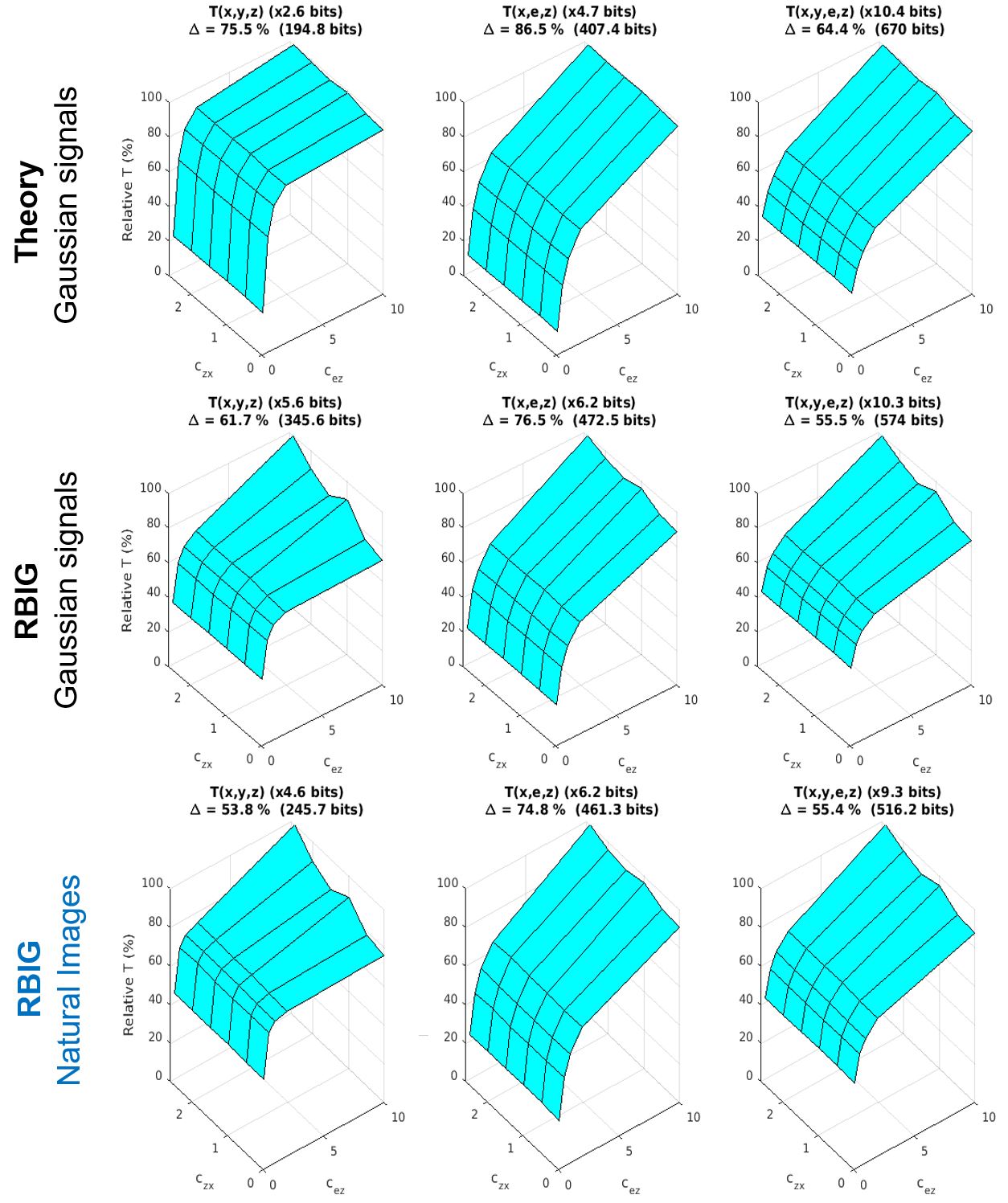}
    \caption{\small{\textbf{Total Correlation strongly depends on the feedforward and feedback connectivity in \emph{Model~II}}. Plots of $T$ as a function of the feedforward connectivity, $c_{ez}$, and feedback, $c_{zx}$, for Gaussian signals (theory and RBIG estimates), and empirical results for natural images.
    The plots display relative values of $T$ in percentage with regard to the maximum together with a factor (e.g. $\times2.6$ in the top-left plot) that allows to express this percentage in absolute values (in bits). Moreover, the plots display the variation (in bits) of the considered descriptor over the range of connectivity values (e.g. $\Delta = 195$ bits in the top-left plot). This is a measure of the sensitivity of the descriptor.}}
    \label{fig:T_feedback}
\end{figure}

Figures~\ref{fig:I_feedback} and~\ref{fig:T_feedback} show the results of Mutual Information and Total Correlation
for different feedforward and feedback connectivity scenarios: different combinations of $c_{ez}$ and $c_{zx}$ in \emph{Model~II}. Specifically, we show:  
(a)~the theoretical results for Gaussian signals,
(b)~the empirical results computed with RBIG for Gaussian signals, and (c)~the empirical results computed with RBIG for natural images. 

In each case the surfaces are plotted in percentage for simpler comparison (flatter surfaces mean less sensitivity and hence worse descriptor). Nevertheless, the numerical captions in each surface give the absolute scale in bits.

%In \emph{Model II}, due to the top-down feedback signal, $\vect{x} \xleftarrow{\,\,c_{zx}\,\,} \vect{z}$, in principle, all the layers can be affected by an enhanced transmission at a deep layer such as, $\vect{e} \xrightarrow{\,\,c_{ez}\,\,} \vect{z}$. Specifically, 
The results for the $I$ and $T$ surfaces show the following major trends:
\begin{enumerate}

     \item The theoretical surfaces in \emph{Model II} are consistently confirmed by the empirical estimations. Similarly to what we found in \emph{Model I}, this parallelism confirms the correctness of the theory and the appropriateness of RBIG in this application.
    
    \item The average percentage of variation of the measures based on $I$ in the theoretical expressions is $\mathbf{\Delta_I = 47 \pm 30 \,\,\,\%}$.

    \item The average percentage of variation of the measures based on $T$ in the theoretical expressions is $\mathbf{\Delta_T = 75 \pm 11 \,\,\,\%}$.

    \item Therefore, the overall sensitivity of $T$ to connectivity and feedback is stronger than the sensitivity of $I$.

    \item Interestingly, the empirical results for natural images also follow the theoretical prediction even though the signals are no longer Gaussian. In this case, the non-Gaussianity only introduces a noticeable variation in  $I(\vect{y},\vect{e})$. However, this does not change much the global sensitivity of $I$, and $T$ is still more sensitive.    

    % 
    % percent I = [30.8 30.9 5.4 49.2 78.8 88.3]
    % percent T = [75.5 86.5 64.4]
    %
    % bits I = [9.7 9.7 9.7 15.4 140.8 288.1]
    % bits T = [194.8 407.4 670]
    %
    % [mean(percentI) std(percentI)]    47   30
    % [mean(percentT) std(percentT)]    75   11
    % [mean(bitsI) std(bitsI)]  79  100
    % [mean(bitsT) std(bitsT)] 400  200
    
\end{enumerate}

%Figure~\ref{fig:T_feedback} shows the results of Total Correlation for different feedforward and feedback connectivity scenarios: different combinations of $c_{ez}$ and $c_{zx}$ in \emph{Model~II}. As above, we show:  
%(a)~the theoretical results for Gaussian signals,
%(b)~the empirical results computed with RBIG for Gaussian signals, and (c)~the empirical results computed with RBIG for natural signals. 
%Here we also present the $T$ surfaces in relative scale for a simpler comparison with the $I$ surfaces in Fig.~\ref{fig:I_feedback}. 

Minor details also include the following: in general, the shared information increases with $c_{ez}$. 
This is obvious in the cases where $\vect{z}$ is one of the considered nodes (e.g. the last three columns $I(\vect{x},\vect{z})$, $I(\vect{y},\vect{z})$ or $I(\vect{e},\vect{z})$) because an increased trasnmission $c_{ez}$ improves the presence of the source in the inner representation. 
More interestingly, we can see that when $\vect{z}$ is not considered, the effect of $c_{ez}$ is only relevant when there is also significant feedback (as in the two first columns $I(\vect{x},\vect{y})$ and $I(\vect{x},\vect{e})$). This is also the case when considering nodes that are far away, as in $I(\vect{x},\vect{z})$.
    
When considering nodes that are far from the considered interactions (e.g. when considering the transmission between $\vect{y}$ and $\vect{e}$ when we consider the forward connection to $\vect{z}$ and the feedback to $\vect{x}$) the mutual information is almost insensitive to the variations of connectivity (see the flat $I(\vect{y},\vect{e})$ in the third column of Fig.~\ref{fig:I_feedback}). 

In summary, the overall sensitivity of $T$ to connectivity and feedback is stronger than the sensitivity of $I$. Note that $\Delta_T > \Delta_I$ with substantially lower variance over the considered nodes. Therefore, $T$ is more appropriate than $I$ to describe the connectivity in the recurrent \emph{Model II}.

\subsection{Results with real fMRI signals from visual regions V1, V2, V3 and V4}
\label{Empiric_results_real}

Once we used sensible analytical scenarios to prove that 
(1) $T$ is more sensitive than $I$ to functional connections up to V1, and 
(2) the empirical estimates through RBIG are reliable for visual signals of dimension $n \in [64, 256]$, 
finally we are ready to use these empirical estimates of connectivity in uncontrolled scenarios down stream.

Here we measure the information shared by different visual regions of the cortex beyond V1. It is true that the previous analytical results (from retina up to V1) give us no direct guarantee of success beyond V1. However the good behavior of the estimates obtained above using signals of similar nature and similar dimension is a necessary safety check which is absent in the purely empirical literature that originally proposed $T$~\cite{Li22,QiangEntr22} or variations~\cite{Herzog22,Gatica21}. 

Measuring $T$ in higher cortical visual areas is interesting because (1) there is a debate on how these regions actually interact~\cite{Semedo21,Kerkoerle14,Klink17,Mejias16,Hulusi15}, and (2) there is a long-standing concept in visual neuroscience that relates neural connectivity with information transmission: the \emph{Efficient Coding Hypothesis}~\cite{Barlow61,Barlow01}.
%,Olshausen96,Schwartz01,Malo06b,Lyu09,Malo10,Gomez19,Malo20}.
Specifically, here we take the neural data from the \emph{Algonauts Project 2021 challenge}~\cite{Cichy21}, and we consider fMRI signals from V1, V2, V3 and V4 while the observers were looking at natural videos.
The details about task paradigm, data acquisition and preprocessing can be seen at http://algonauts.csail.mit.edu/2021/brainmappingandanalysis.html. 
% are given in Appendix D.
In our experiments we consider pairwise and multivariate relations among regions which (anatomically) are progressively farther away. However, our descriptors of functional links do not make any prior assumption of the possible feedforward or feedback connections.

\textbf{Ensembles:} The considered dataset provides 3 responses of 9 observers for 1000 natural videos in a number of voxels of the considered regions (V1, V2, V3 and V4). 
In this database there is a one-to-one relation between input and responses, but the number of available voxels depends on the observer and the cortical region. Therefore, just for illustrative purposes, we take 20 randomly selected voxels per region for each observer. This means 20-dimensional signals associated to one input.
By considering the data of all trials, all observers, and all input videos, we have $3\times9\times1000 = 27000$ samples of these 20-dimensional vectors for each region. In these ensembles, the i-th vector of each region corresponds to the same input and the same observer, but the j-th dimension of the vector is the response of a randomly chosen voxel in that region (and observer). 
% so the meaning of the j-th dimension is not uniform accross the 27000 samples: 
We assume all the observers and all the voxels in a region are equivalent.
By rerunning this random selection of voxels we get equivalent ensembles.

\textbf{Empirical estimation with fMRI data using RBIG:} Given the fact that the marginal PDFs of the considered fMRI signals are approximately Gaussian (results not shown), in the estimations of $T$ and $I$ based on iterative Gaussianization we chose a small number of iterations (only 20 iterations as opposed to the 500 iterations used in \emph{Model I} where $\emph{z}$ is non-Gaussian). 
We re-estimate $T$ and $I$ 30 times from equivalent, randomly chosen, ensembles, and we report the average and standard deviation of the results. 

\begin{figure}[t!]
    \begin{centering}
    \hspace{-0cm}\includegraphics[width=1.0\textwidth]{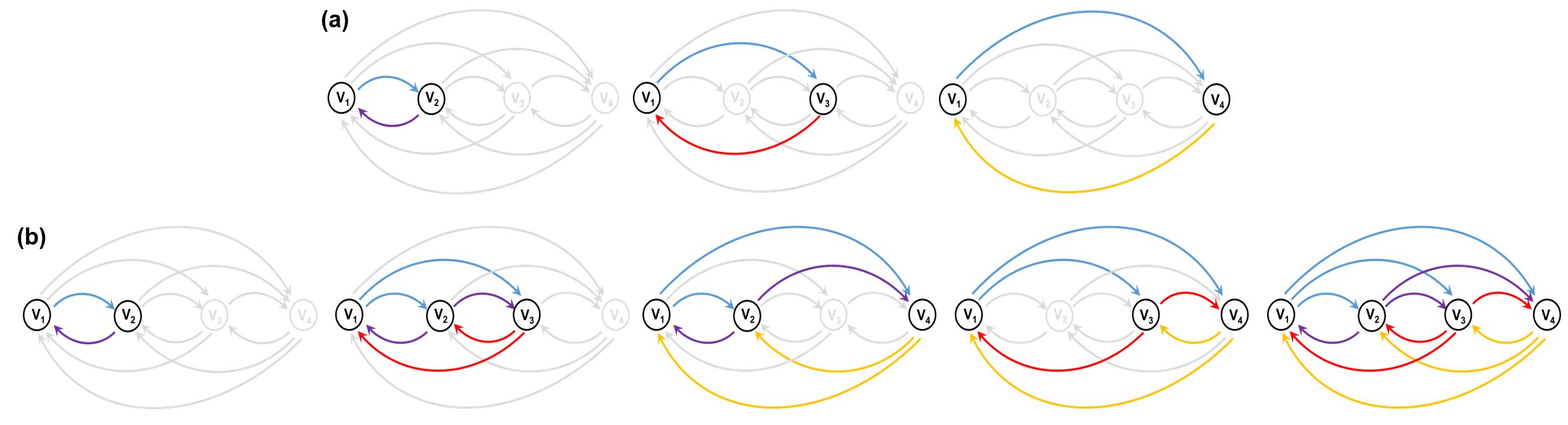}
    \caption{\small{\textbf{Examples of the pairs of nodes (a), and groups of nodes (b), that we consider in our measurements}. The top row (a) considers the relation between certain node (in this case V1), and, from left to right, nodes progressively distant: V2, V3, and V4. The bottom row (b), from left to right, adds nodes to the group under consideration. If we start from $V_i$, the added nodes can be close to it, e.g. $(V_{i}, V_{i+1}, V_{i+2})$ in the 2nd diagram, or they can be progressively farther away, as $(V_i, V_{i+1}, V_{i+3})$ or $(V_i, V_{i+2}, V_{i+3})$, as in the 3rd and 4th diagrams. Finally, the last diagram at the right shows that we can consider all nodes at the same time, namely $(V_1, V_2, V_3, V_4)$.}}
    \label{fig:possibilities_real_data}
    \end{centering}
\end{figure}

\textbf{Measurements of functional links:} we estimated $I$ and $T$ in all possible distinct combinations of nodes. Figure~\ref{fig:possibilities_real_data} illustrates pairwise and multivariate relations among regions which (anatomically) are progressively farther away.
Note that the functional link of the configurations in the top row can be addressed by the pairwise $I(\vect{v_i},\vect{v_j})$ or $T(\vect{v_i},\vect{v_j})$. However, progressive consideration of additional nodes, as in the bottom row, can only be quantified using a multivariate descriptor
$T(\vect{v_i},\vect{v_j},\vect{v_k},\ldots)$.
Note that in a case where the connections are unknown, the shared information (either $I$ or $T$) is not only affected by the \emph{direct} connections between the considered nodes (in our figure \emph{direct} connections are in color), but also by all other possible \emph{indirect} connections (depicted in gray). 
The \emph{indirect} connections imply communication through alternative regions that may re-inject the relevant signal into the considered nodes and have a positive effect in the functional link.

%\textbf{Information in a single node as reference:} 
On top of the two-node and multi-node cases, mono-mode references are convenient to know if the information is lost through the network or, on the contrary, there are positive synergies. 
To this end, we report three additional numbers: $T(\vect{v_i})$, which is a measure of the redundancy within the node $\vect{v_i}$; and also $I(\vect{v_i},\vect{v_i})$, and $T(\vect{v_i},\vect{v_i})$. 
In principle, the information shared by a variable with itself, as in $I(\vect{v_i},\vect{v_i})$, and $T(\vect{v_i},\vect{v_i})$, is $\infty$~\footnote{Given any $n$-dimensional variable $\vect{a}$, the samples of $(\vect{a},\vect{a})$ are aligned in a $2n$-dimensional space, and then the joint differential entropy terms of Eqs.~\ref{defT}-\ref{defI} is $-\infty$, leading to $I(\vect{a},\vect{a})=T(\vect{a},\vect{a})=\infty$.}. However, given the uncertainty we introduce when using random voxels from each region/observer, two (randomly chosen) sets of $\vect{v_i}$ are not aligned and then $I(\vect{v_i},\vect{v_i})$, and $T(\vect{v_i},\vect{v_i})$ do not diverge to $\infty$. Instead, they are measures of the common information present in every realization of the ensemble of responses of that node $\vect{v_i}$. Therefore, they are a convenient reference to know if the consideration of extra nodes increases or decreases this mono-mode amount of information.
%In this regard, $I(\vect{v_i},\vect{v_i})$ does not take the redundancy between voxels into account while $T(\vect{v_i},\vect{v_i})$ does consider it.

For a more intuitive comparison of the results corresponding to configurations with different number of nodes, we report the shared information \emph{per node}. This means: $I(\vect{v_i},\vect{v_j})/2$, $T(\vect{v_i},\vect{v_j})/2$, $T(\vect{v_i},\vect{v_j},\vect{v_k})/3$, and, $T(\vect{v_1},\vect{v_2},\vect{v_3},\vect{v_4})/4$.
In the case of $T(\vect{v_i})$ the definition already has a single node, so \emph{bits} and \emph{bits/node} are the same.

Finally, we report not only the absolute values in \emph{bits/node}, but (more interestingly to describe the connectivity) how the information per node increases or decreases when we go way from one node or include progressively distant nodes in the measure. We give this deviation in $\%$ with regard to the information per node in V1 (either $I(\vect{v_1},\vect{v_1})$ or $T(\vect{v_1},\vect{v_1})$).

\textbf{Results:} Tables~\ref{table1}-\ref{table2} show the measures 
of shared information in three panels: the top panel shows the pair-wise measures $I(\vect{v_i},\vect{v_j})$, the middle panel shows the single-node measure $T(\vect{v_i})$, and the bottom panel shows the multi-node measures $T(\vect{v_i},\vect{v_j},\ldots)$.
Table~\ref{table1} has absolute measures in \emph{bits/node}, and Table~\ref{table2} displays the variation (in $\%$) of the considered configuration with regard to the corresponding measure in V1.
The $T(\vect{v_i},\vect{v_j},\ldots)$ panels have a pair-wise part (at the left) and a multi-node part (the last four columns). This multi-node parts have to be read \emph{row-wise}: each number reports how the node in the row interacts with the nodes in the different columns. Moreover, the consideration of extra nodes is done in cyclic way: in the 3rd row $v_i=v_3$, and hence the 5th column, $(v_{i+1},v_{i+2})=(v_4,v_1)$, refers to the connectivity among the nodes $(v_3,v_4,v_1)$.

\begin{table}[b!]
\begin{centering}
\begin{tabular}{|c||c|c|c|c|}
	\hline
	$\mathbf{I(v_i,v_j)}$ & & & & \\
	(in \emph{bits/node})  & $v_1$ \  & $v_2$ & $v_3$ & $v_4$ \\
	\hline\hline
	$v_1$ & $\blau{\textbf{2.3}} \pm0.3$ & $\blau{\textbf{1.4}} \pm0.4$ & $\blau{\textbf{1.0}} \pm0.2$ & $\blau{\textbf{0.7}} \pm 0.2$ \\
	\hline
	$v_2$ & {1.4}  & $\blau{\textbf{2.0}} \pm0.3$ & $\blau{\textbf{1.3}} \pm0.2$ & $\blau{\textbf{0.7}} \pm0.1$\\
	\hline
	$v_3$ & ${1.0}$  & ${1.3}$ & $\blau{\textbf{1.7}} \pm0.3$ & $\blau{\textbf{0.8}} \pm0.2$\\
	\hline
	$v_4$ & ${0.7}$  & ${0.7}$ & ${0.8}$ & $\blau{\textbf{2.2}} \pm0.3$ \\
	\hline
\end{tabular}
%\end{centering}
\\\vspace{0.4cm}

\begin{tabular}{|c||c|c|c|c|}
	\hline
	$\mathbf{T(v_i)}$ & & & & \\
	(in \emph{bits/node})  & $v_1$ \  & $v_2$ & $v_3$ & $v_4$ \\
	\hline\hline
	           &$\blau{\textbf{3.5}} \pm0.3$ & $\blau{\textbf{3.2}} \pm0.3$ & $\blau{\textbf{3.0}} \pm0.3$ & $\blau{\textbf{3.4}} \pm 0.2$ \\
	\hline
\end{tabular}
%\end{centering}
\\\vspace{0.4cm}

%\begin{centering}
%\hspace{-1cm}
\begin{tabular}{|c||c|c|c|c||c|c|c||c|}
	\hline
	${\mathbf{T(v_i,v_j,\ldots)}}$  & & & & & & & & \\
	{(in \emph{bits/node})} & $v_1$ & $v_2$ & $v_3$ & $v_4$ & $v_{i+1},v_{i+2}$ & $v_{i+1},v_{i+3}$ & $v_{i+2},v_{i+3}$ & $v_{i+1},v_{i+2},v_{i+3}$ \\
	\hline\hline
	$v_1$ & $\blau{\textbf{6.0}} \pm0.3$ & $\blau{\textbf{5.1}} \pm0.3$ & $\blau{\textbf{4.7}} \pm0.3$ & $\blau{\textbf{4.6}} \pm0.2$ & $\blau{\textbf{6.1}} \pm0.3$ & $\blau{\textbf{5.9}} \pm0.3$ & $\blau{\textbf{5.7}} \pm0.3$ & $\blau{\textbf{6.6}} \pm0.3$ \\
	\hline
	$v_2$ & ${5.1}$  & $\blau{\textbf{5.4}} \pm0.3$ & $\blau{\textbf{4.7}} \pm0.3$ & $\blau{\textbf{4.5}} \pm0.3$ &  $\blau{\textbf{5.7}} \pm0.3$ & $6.1$ & $5.9$ & $6.6$ \\
	\hline
	$v_3$ & ${4.7}$  & ${4.7}$ & $\blau{\textbf{5.0}} \pm0.3$ & $\blau{\textbf{4.5}} \pm0.3$ & $5.7$ & $5.7$ & $6.1$ & $6.6$ \\
	\hline
	$v_4$ & $4.6$  & $4.5$ & $4.5$ & $\blau{\textbf{5.9}} \pm0.3$ & $5.9$ & $5.7$ & $5.7$ & $6.6$ \\
	\hline
\end{tabular}
\vspace{0.2cm}
\caption{\blue{\small{\textbf{$I(\vect{v_i},\vect{v_j})$ between pairs of areas, 
$T(\vect{v_i})$ in each area, and $T(\vect{v_i},\vect{v_j},\ldots)$ among multiple areas in \emph{bits/node}}. 
The reported values are the mean and the standard deviation of the corresponding magnitudes over 500 estimations using independent datasets. The independent configurations are highlighted in \emph{blue}. The non-highlighted values correspond to symmetry-equivalent configurations.
See the \textbf{\emph{Results}} paragraph in the text for the interpretation of pairs and triplets with progressively distant nodes.}}}
\label{table1}
\end{centering}
\end{table}

Not all the values in the tables are independent because of the symmetry of the measures.
Note that $I$ and $T$ are invariant to the permutation of the variables: $I(\vect{v_i},\vect{v_j}) = I(\vect{v_j},\vect{v_i})$, and $T(\vect{v_i},\vect{v_j},\vect{v_k}) = T(\vect{v_j},\vect{v_k},\vect{v_i}) = \ldots $
This implies that the $I$ panels are symmetric and so it is the pairwise part of the $T$ panels. 
Also as a consequence of the invariance to permutation, some multi-node configurations are equivalent. As the order does not matter, we have combinations of 4 nodes taken 3 at a time, i.e. only 4 independent node configurations. 
For the sake of clarity the non-redundant values of the tables are highlighted in \emph{blue}.  
\blue{Also for clarity, the standard deviation over the 500 realizations of the estimation has been reported only in the independent values of Table~\ref{table1}}.

The discussion of the results will be focused on the variations of information as we depart from a node (Table~\ref{table2}). Departure, as in the top row of Fig.~\ref{fig:possibilities_real_data}, means \emph{moving away from the diagonal} (along rows/columns) in the pairwise parts of the tables. 
Departure, as in the bottom row of Fig.~\ref{fig:possibilities_real_data}, means \emph{moving to the right (for the highlighted numbers)} in the multi-node parts. Table~\ref{table1}, with the original absolute measures, is just given for completeness and for the reader convenience. 

A final comment on the absolute magnitudes: in every case, the estimated $T(\vect{v_i},\vect{v_j}) > I(\vect{v_i},\vect{v_j})$, which is consistent with the definitions because (as discussed in Eqs.~\ref{defT}-\ref{defI}, \blue{and in Appendix C}) $T$ includes the redundancy within the nodes and hence the information is necessarily bigger.

\begin{table}[t!]
\begin{centering}
\begin{tabular}{|c||c|c|c|c|}
	\hline
	$\mathbf{\Delta I(v_i,v_j)}$ & & & & \\
	(in $\%$)  & $v_1$ \  & $v_2$ & $v_3$ & $v_4$ \\
	\hline\hline
	$v_1$ & \blau{\textbf{0}} & \blau{\textbf{-41}} & \blau{\textbf{-56}} & \blau{\textbf{-69}} \\
	\hline
	$v_2$ & -41 & \blau{\textbf{-14}} & \blau{\textbf{-46}} & \blau{\textbf{-69}}\\
	\hline
	$v_3$ & -56 & -46 & \blau{\textbf{-27}} & \blau{\textbf{-66}}\\
	\hline
	$v_4$ & -69 & -69 & -66 & \blau{\textbf{-7}} \\
	\hline
\end{tabular}
\\\vspace{0.4cm}

\begin{tabular}{|c||c|c|c|c|}
	\hline
	$\mathbf{\Delta T(v_i)}$ & & & & \\
	(in $\%$)  & $v_1$ \  & $v_2$ & $v_3$ & $v_4$ \\
	\hline\hline
	           & \blau{\textbf{0}} & \blau{\textbf{-8}} & \blau{\textbf{-15}} & \blau{\textbf{-1}} \\
	\hline
\end{tabular}
%\end{centering}
\\\vspace{0.4cm}
\resizebox{\linewidth}{!}{%
\begin{tabular}{|c||c|c|c|c||c|c|c||c|}
	\hline
	$\mathbf{\Delta T(v_i,v_j,\ldots)}$  & & & & & & & & \\ 
	(in $\%$) & $v_1$ & $v_2$ & $v_3$ & $v_4$ & $v_{i+1},v_{i+2}$ & $v_{i+1},v_{i+3}$ & $v_{i+2},v_{i+3}$ & $v_{i+1},v_{i+2},v_{i+3}$ \\
	\hline\hline

	$v_1$ & \blau{\textbf{0}} & \blau{\textbf{-15}}  & \blau{\textbf{-21}} & \blau{\textbf{-23}}&                                                        
    \blau{\textbf{2}} & \blau{\textbf{-1}} & \blau{\textbf{-4}} & \blau{\textbf{11}} \\
	
	\hline
	$v_2$ & -15 & \blau{\textbf{-9}} & \blau{\textbf{-20}} & \blau{\textbf{-25}} & 
	\blau{\textbf{-5}} & 2 & -1 & 11 \\
	
	\hline
	$v_3$ & -21  & -20 & \blau{\textbf{-16}} & \blau{\textbf{-25}} &
	 -4 & -5 & 2 & 11 \\
	
	\hline
	$v_4$ & -23  & -25 & -25 & \blau{\textbf{-1}} &
	 -1 & -4 & -5 & 11 \\
	\hline
\end{tabular}%
}
\vspace{0.2cm}
\caption{\blue{\small{\textbf{Variations of $I$ and $T$ (in \% with regard to V1) when considering progressively distant nodes or extra nodes}.
Negative numbers imply information loss and positive increments indicate a sort of synergy.
See the \textbf{\emph{Results}} paragraph in the text for the interpretation of pairs and triplets with progressively distant nodes.}}}
\label{table2}
\end{centering}
\end{table}

\textbf{Information flow and conjectures on connectivity:}
Results show that the redundancy within each node $T(\vect{v_i})$ is smaller in deeper layers than in V1 (see the negative increments in the middle panel of Table~\ref{table2}). This is consistent with the \emph{Efficient Coding Hypothesis}~\cite{Barlow61,Barlow01}.

Reduction in $T(\vect{v_i})$ in the middle panel is not the same as the reductions of $T(\vect{v_i},\vect{v_i})$ or $I(\vect{v_i},\vect{v_i})$ along the diagonal of the pairwise parts of the top and the bottom panels. While redundancy reduction in $T(\vect{v_i})$ means better information encoding, reduction in $T(\vect{v_i},\vect{v_i})$ or $I(\vect{v_i},\vect{v_i})$ means a decay in the information content. 
This decay is more apparent in $I(\vect{v_i},\vect{v_i})$, because the reduction of $T(\vect{v_i},\vect{v_i})$ is biased by the simultaneous reduction of the intra-node redundancy in $T(\vect{v_i})$. Actually, if we discount $T(\vect{v_i})$ from $T(\vect{v_i},\vect{v_i})$, the corrected variations may change their sign and become a gain\footnote{Variations $\Delta T$ corrected in this way (remaining information after discounting redundancy) are -11$\%$, -17$\%$, and +1$\%$, for v2, v3, and v4, respectively. This implies an increment in V4, while with the original values one gets (-9$\%$, -16$\%$, -1$\%$), as shown in the diagonal of the pairwise part of the $T$ panel (in Table~\ref{table2}). This positive variation is in contrast with the $-7\%$ loss in $\Delta I(v4,v4)$.}. These kind of gains may be a positive effect of connectivity seen in $T$ and not in $I$.

However, the mono-node measures mentioned above only describe the information in each node, but not how much of this information comes from another region. This second concept, more related to connectivity, is measured by pairwise and multi-node measures.
In this regard, progressively bigger reductions in the pairwise $\Delta I(\vect{v_i},\vect{v_j})$ and $\Delta T(\vect{v_i},\vect{v_j})$ away from the diagonal mean information loss along the way (or reduced functional connectivity).
This information loss seems consistent with the \emph{data processing inequality}~\cite{Cover06} to a certain extent. However, as discussed below, the results (particularly $T$ in multiple nodes) confirm the existence of relevant feedback in these regions. 

The \emph{data processing inequality}~\cite{Cover06} states that information lost between two nodes cannot be recovered by further processing (with no additional input from the original node). This inequality strictly holds in purely feedforward schemes $\vect{v_1}\rightarrow\vect{v_2}\rightarrow\vect{v_3}\rightarrow\vect{v_4}$, where, due to the absence of feedback connections and skip connections, the response in inner layers conditioned to the previous layer is independent of the early layers. In such systems, it holds $I(\vect{v_1},\vect{v_2})>I(\vect{v_1},\vect{v_3})>I(\vect{v_1},\vect{v_4})$.   
This behavior is what is observed in the rows of the $I$ panel when moving away from the diagonal to inner layers. This suggests that the feedforward component of the connectivity can be strong, and in such simplistic situation, one could deduce the strength of each connection from the different decays in $I(\vect{v_i},\vect{v_j})$. 

However, in our case (where feedback and skip connections may exist) the \emph{data processing inequality} may not hold. 
Reductions in $I$ do not necessarily mean that the other connections are not present. This is more clear looking at the results of $T$. While the behavior of the pairwise $T$ moving to deeper layers is negative (similarly to $I$), something different happens by  considering extra nodes. 
Under the purely feedforward assumption extra nodes should share less information with the previous and the global $T$ should decrease, particularly if the intra-node redundancy does not increase (as in this pathway). However, we see that in some cases the consideration of extra nodes implies an increase of the shared information per node, as for instance when going from $(v_1,v_2)$ to $(v_1,v_2,v_3)$ or from there to $(v_1,v_2,v_3,v_4)$ (see the positive increments highlighted in blue in Table~\ref{table2}). 

Multi-node results obtained from the proposed measure $T$ are interesting because we can see that the connections in the group $(v_1,v_2,v_3)$ are a bit stronger than the connections in the group $(v_2,v_3,v_4)$ despite they are at similar anatomical distance. This suggests some top-down feedback from $v_3$ or $v_2$ or feedforward skip connections from $v_1$ to $v_3$. The same is true when considering all the nodes together with a substantial increment (by 11$\%$). See the raw data in Appendix D (histograms) to see the differences in the values.

These two different synergistic behaviors that can be seen using the proposed \emph{Total Correlation} clearly mean that 
one can rule out a pure feedforward scheme in the $V_1 ,V_2,V_3,V_4$ regions, and more complex connectivity schemes do exist.
This is not that obvious just using the conventional $I$.

\section{Discussion and conclusions}
\label{Discussion}

\textbf{Analytical results: $T$ is a better descriptor of connectivity than $I$.}
The goal of this paper is addressing the fundamental limitation of the seminal work that proposed $T$ as a measure of functional connectivity~\cite{Li22}: namely the lack of analytical results that can justify the superiority of the $T$ over the conventional $I$ beyond the multivariate versus pairwise definitions. Here we did that analytical study in the context of the early visual brain with simple models of the retina-V1 cortex pathway. 

For mathematical convenience we considered two variations of the general framework presented in the \blaveros{diagram}~\ref{Framework}: \emph{Model~I} and \emph{Model~II}. 
These models were chosen to illustrate two fundamental properties of neural architectures in early vision: (1)~the~Divisive Normalization nonlinearity in \emph{Model~I}, in Section~\ref{Models_no_recurrence}, and (2)~an eventual top-down recurrence in \emph{Model~II} in Section~\ref{Models_recurrence}.

It is important to stress again that the models are not arbitrary: according to the results in Section~\ref{Models_plausibility} the nonlinearity in \emph{Model~I} is key to improve the explanation of the psychophysics, and the explored range of intra-cortical connectivity actually covers different behaviors (with substantial differences in the explained variance of human data). The top-down connection in \emph{Model~II} was not specifically justified, but given the observed behavior of the steady state in $\vect{e}$, the explored feedback does not reduce substantially the $\rho = 0.7$ result. This indicates that \emph{Model~II} has certain biological plausibility, so that it can be used to illustrate the study of recurrent connections.  
The plausibility of the models and the generality and relevance of the facts they illustrate (nonlinearities and recurrence) implies that a proper descriptor of functional connectivity should be sensitive to the different variations of the models. 

Sections~\ref{res_anal_1}, \ref{Anal_emp_results_no_rec},
and~\ref{Anal_emp_results_rec} explicitly show the superiority of $T$ over $I$ in the considered nonlinear and recurrent models. 
The conclusion of these analytical results (confirmed by the  experimental simulations) is that while the conventional \emph{Mutual Information} is not useful to capture the intra-cortical connections in \emph{Model~I}, the proposed measure, \emph{Total Correlation}, is quite sensitive to this connectivity.
Similarly, the proposed \emph{Total Correlation} is more sensitive than \emph{Mutual Information} to the feedforward and feedback connectivity explored in the recurrent \emph{Model~II}.
From a general perspective, the considered nonlinearity is ubiquitous in the visual pathway~\cite{Carandini12,Brainard05,Watson97,Simoncelli98,Martinez18}. Therefore, the success of the proposed multivariate \emph{Total Correlation} in describing this connectivity is a substantial advantage with regard to the conventional, pairwise, \emph{Mutual Information}.

\textbf{\blaveros{Temporal dynamics can be incorporated in the theory.}} 
\blaveros{Transmission time and recurrence implied by feedback imply a nontrivial evolution of the signals when the system faces dynamic inputs with fast variations compared with the updating time constant $\Delta t$. 
In our simulations we consider slow-varying sources $\mathbf{s}(t)$ and (in Model II) we wait till the convergence of the signals to a stationary state to measure the statistical dependence between the signals at the different layers. That situation is equivalent to assuming static signals (corresponding to the stationary situation) and zero communication delay between layers.
The consideration of the biophysics of communication and the resulting delay may certainly modify the proper (best corresponding moment) to look for maximum relations.}

\blaveros{However, these assumptions are not a major restriction of the results. This is because the fundamental properties invoked to prove the superiority of $T$ are time independent. Therefore, the proposed $T$ is still expected to be more sensitive to changes in connectivity than the traditional pairwise $I$ even if time delays are different from zero.}

\blaveros{Looking at the proposed analytical expressions~\ref{Eq.deltaf},~\ref{one_II} and ~\ref{two_II},~\ref{more_II}, delays just impact on the expected values that define the covariance matrices involved in $T$ and $I$.
Let's consider the effects in turn, first in the nonlinear model, and then in the model with feedback.}

\blaveros{In the nonlinear model delayed transmission does not affect the diagonal blocks of the covariance matrices because they describe interaction between the signal within certain layer (and hence at a fixed time). Only the off-diagonal blocks are affected because they consider the relation between the signal at different layers (and, given the transmission delay, at different times). Therefore, tracing the signals at $\vect{y}$, $\vect{e}$, or $\vect{z}$, back to the signal at $\vect{x}$, in the covariance one would have comparisons between the values of $\vect{x}$ at different times, which certainly would imply a modification: $\Sigma_x = E[x(t) x(t)^\top] \neq \Sigma’_x = E[x(t) x(t-\Delta t)^\top]$. The modification may be due to two reasons: (1) if the stimulus $\vect{s}$ is stationary, correlations may be reduced because of ocular motion and may be increased because of averaging independent realizations of the noise in the photoreceptors. (2) if the stimulus $\vect{s}$ is not stationary (as in video sequences), correlation between the signal values at different locations will be also decreased due to motion in the scene. However, assuming an auto-regressive model for natural videos (which is a sensible rough model that allows robust motion estimation in video coding~\cite{Tekalp15, Malo01}, and justifies spatio-temporal DCT-like eigen functions for natural video~\cite{Laparra15}) one could propose an expression for these $\Sigma’_x$ or even compute them empirically from samples. In any case, note that these modifications do not change the analytical result because transmission delays in the covariance matrices in Eqs.~\ref{Eq.deltaf},~\ref{one_II}, would modify the specific values of $T$ and $I$, but do not modify the fact that $f(\cdot)$ depends on the intra cortical connectivity, and hence $T$ is sensitive to that connectivity while $I$ is not.}

 \blaveros{For the model with feedback: the reasoning for the off-diagonal terms in Eqs.~\ref{two_II} and~\ref{more_II} is exactly the same as the one given above for the covariances in Eqs.~\ref{Eq.deltaf},~\ref{one_II}. However, with feedback the diagonal blocks also change because they imply comparison between delayed signals $\vect{x}$ at different times. Specific simulations could determine how these variations will impact on the determinants involved in the entropies of Gaussians in Eqs.~\ref{defI} and~\ref{variaT}, but it is important to stress that: (1) the described effect is the same in the expressions for $I$  and $T$
and, more importantly, (2) all the proposed expressions (and code provided) are valid with the corresponding modification of the covariance $\Sigma’_x$. so the theory could be used to repeat the computations with empirical estimations of this covariance.}

\blaveros{The theory (expressions and code) could be used to explore different choices of delay in situations before the stationary state has been reached, or with non-stationary stimuli. 
Similarly, temporal variations of connectivity due to adaptation (e.g. changes in the Divisive Normalization kernel in different environments~\cite{Coen12}) could be studied with the proposed theory.
However, a detailed analysis (with a variety of options for the delay and feedback factors) is a separate research which is out of the scope of this work and matter of future research.}

% In summary, there is no problem in using $T$ for signals acquired at different delays. In that situation, one may define the functional connectivity for the delay that maximizes the proposed measure, or (for fair comparison with previous studies) take the delay that is assumed with traditional pairwise measures.

\textbf{\blue{Introducing certain sensitivity to direction}.} 
\blue{Classical measures as the correlation, and also the measures compared here ($T$ and $I$) are not directional. As a result, the forward/feedback possibilities are, by definition, not easily distinguishable. 
However, when multiple nodes are considered ($T$ can consider many at once, which cannot be done by $I$ nor by the correlation) 
one could look at variations in the amount of \emph{information per node}. Modifications of $T$/node when including extra nodes in certain order may give insights on the direction of the relations.}

\blue{Additionally, one could introduce some sort of sensitivity to direction in $T$ \emph{by conditioning} as done with mutual information in transfer entropy~\cite{Schreiber00}. By applying the information-chain-rule (as in transfer entropy) one could also reduce the problem to the estimation of joint entropy values.} 
%which could be computed either analytically (in simple cases –e.g. Gaussian-), or, more generally, using reliable empirical estimators.

\textbf{Results with real data: $T$ highlights synergies in $V_1, V_2,V_3,V_4$.} The positive results of $T$ (and the corresponding RBIG estimates) in the analytical settings presented above not only address a limitation of~\cite{Li22,QiangEntr22}, but really justify its use in real scenarios.
In the case of fMRI data from the visual regions 
$V_1,V_2,V_3,V_4$, our measurements of $T$ show that: (1) the redundancy within each layer, $T(v_i)$, is reduced along the way, which is consistent with the \emph{Efficient Coding Hypothesis}, (2) the information content measured through $T(v_i,v_i)$ is more stable along the way than the measures given by $I(v_i,v_i)$, particularly if the inner redundancy is discounted. (3) The variation of the pairwise measures of $I(v_i,v_j)$ seems compatible with the \emph{data processing inequality} in a purely feedforward setting $\vect{v_1}\rightarrow\vect{v_2}\rightarrow\vect{v_3}\rightarrow\vect{v_4}$,
however, (4) the multi-node $T$ shows synergies that rule out the purely feedforward scheme. Moreover, it suggests stronger functional connectivity between the nodes $V_1,V_2,V_3$ than between $V_2,V_3,V_4$ despite a similar anatomical distance.
All this complex behavior is not easy to see just using the conventional $I$.

\textbf{Relations with previous work.}
Firstly, this is the necessary analytical companion of the proposal of \emph{Total Correlation} to measure connectivity~\cite{Li22,QiangEntr22}.
Then, here we have applied this tool to visual areas extending the works that first used Mutual Information to assess the connectivity between pairs of visual areas~\cite{FeiFei09} or those that measured Mutual Information between V1 and MT (or V5) under Divisive Normalization transforms~\cite{Serences14}. The analysis of Mutual Information between progressively deeper visual layers is also related with previous works focused on quantifying the information flow in different nonlinear models of retina-V1 pathway~\cite{Gomez19,Malo20}, which were restricted to purely feedforward models.

On the other hand, the approach we took here (quantifying the statistical properties of the responses of real brains or psychophysically plausible models) is 
is related with a body of literature that follows Barlow's \emph{Efficient Coding Hypothesis} in a non-classical direction.
Note that the classical direction is \emph{from-statistics-to-biology}: a system optimized for a sensible statistical goal may display biological-like behavior~\cite{Barlow61,Barlow01}.
This is the direction that explained linear receptive fields~\cite{Atick92,Li22csfs,Field96,Gutmann14} and sensory nonlinearities~\cite{Schwartz01,Twer01,Malo06b,Lyu09,Laparra12,Laparra15} from statistics.
However, there is literature that reasons in the opposite direction \emph{from-biology-to-statistics}: look at the statistical properties of the responses of biologically plausible systems and you will find statistically interesting behavior. In this regard, redundancy reduction~\cite{Parga10,malo06a,Malo10}, and efficient information transmission~\cite{FosterCIC08,Gomez19,Malo20,Malo2019FI} has been found in in real and biologically plausible models. And this is similar to the information-theoretic analysis that we did of real and simulated responses.

\textbf{Limitations and future research.} 
This study has different limitations that should be addressed by future research. 
First (and most important) is the unification of the analytical examples: they addressed fundamental issues such as nonlinearities and recurrence, but they did it in \emph{separate} examples (\emph{Model I} and \emph{Model II}). Moreover,
\emph{Model I} didn't include noise after the divisive normalization so that one could apply the property of the variation of $T$ under deterministic transforms, Eq.~\ref{variaT}, and the invariance of $I$ under transforms of one of the variables, Eq.~\ref{invarianceI}. Future research should try to get unified expressions for a general nonlinear and recurrent model with noise at all layers.

Second, we left out the comparison with other interesting pairwise (but directional) measures related to Mutual Information such as Transfer Entropy and Granger Causality~\cite{Barnett09}. 
As mentioned above, this would require extensions of $T$ by conditioning on the past values of the signals. \blaveros{Regarding other linear measures like coherence~\cite{Sun04}, partial cross-correlation~\cite{PartialCrossCorrelation05}, or phase synchrony~\cite{PhaseLock99}, we just present one empirical illustration of their behavior in Appendix~E that suggest their inability to capture connectivity in nonlinear settings (with the default implementation in~\cite{Zhou09}). Nevertheless, a detailed theoretical account of this inability is out of the scope of this work.}

The third, more instrumental, limitation is related to the specific empirical estimator of $T$ which is necessary in real scenarios. Here we used our \emph{Rotation-Based Iterative Gaussianization}~\cite{Laparra11,Laparra20}, and it proved to follow the trends of the theoretical surfaces in the analytical scenarios. However, RBIG may suffer from errors when the signals are strongly non-Gaussian with multiple modes separated by low probability regions as may happen after Divisive Normalization (see the PDFs of natural images in~\cite{Malo10,Malo20}). An approximate knowledge of the PDF of the signals is required to set the number of iterations in RBIG. Of course, future research can use other empirical estimators as for instance~\cite{szabo14,Marin-Franch13,steeg2014NIPS,steeg17,steeg2015corex_theory}.
In this regard, the analytical results presented here are a good test-bed for current or future empirical estimators. 

Finally, regarding the results with real data, it is important to acknowledge that there are more comprehensive databases. The one we used (\emph{the Algonauts 2021 Challenge}~\cite{Cichy21}) only considers 1000 videos and has a restricted set of voxels because we wanted a simple proof of concept for our measure $T$ and estimator RBIG on low level regions.
The work done here could be extended in different ways. First, the database could be segmented depending on the properties of the stimuli (e.g. color, texture and motion content) because the functional connectivity between the considered regions may depend on these low-level features of the input. 
This could tell us about the specialization of these regions in different dimensions of the stimuli.
Moreover, the computation of connectivity based on $T$ depending on the structure of the scene could clarify the differences in the feedback signals found in figure-ground contexts~\cite{Klink17,Hulusi15}.
And second, larger databases (such as~\cite{NaturalScenes_21}) may be convenient to confirm the current results and be more appropriate to study the connectivity depending on the properties of the input so that the subsets are big enough to trust the information estimates. Databases like~\cite{BOLD5000_19} can be used to address the relation between V1 and higher-level regions (FFA, PPA,...).

%* \textcolor{blue}{QL:Add appendix-see information in our previous prepared PPT}.

\textbf{Conclusions:} 
In this work we derived analytical results that show that \emph{Total Correlation} is a better descriptor of connectivity than \emph{Mutual Information} in plausible models of the retina-LGN-V1 that include nonlinearities due to intra-cortical connectivity and top-down feedback. $T$ is better because it is more sensitive than $I$ to connectivity.
Analytical results are derived for Gaussian signals but, as confirmed by empirical estimates, they also hold for natural inputs. Our $T$ results for real responses recorded from V1,V2,V3,V4 rule out a naive feedforward-only information flow and suggest stronger feedback connections in V1,V2,V3, than in V2,V3,V4.

The proposed measure opens several possibilities: (1) it can be applied to assess the connectivity in complex models that have been developed to reproduce feedforward and feedback oscillations~\cite{Mejias16}, and (2) it can be used to examine signal-dependent feedback in stimuli with figure-ground or spatially segregated textures, which is an interesting open question in visual neuroscience~\cite{Klink17,Hulusi15}. 

%\section{Author contributions}
%\label{Auth}
%\textbf{QL} checked the plausibility of the neural models reproducing the psychophysical data, implemented the experiments to compare Total Correlation and Mutual Information,  computed the measures in the fMRI data using RBIG and contributed to the writing of the manuscript.
%\textbf{GVS} made some comments and suggestions on earlier versions of this manuscript, and contributed to the writing of the manuscript.
%\textbf{JM} (corresponding author) had the idea of the work, developed the theoretical results, implemented the neural model, and prepared the first draft of the manuscript. 

\section{Acknowledgements}
\label{Acks}
The authors thank Dr. Valero Laparra for his comments on early stopping of RBIG depending on data dimensionality and Dr. Olga Stefanska for her support in the writing process.
We thank the organizers of the \emph{Algonauts Project 2021 Challenge} for providing their interesting fMRI dataset which was used in this study.
This work was partially funded by these spanish/european grants from GVA/AEI/FEDER/EU:  MICINN PID2020-118071GB-I00, MICINN PDC2021-121522-C21, and GVA Grisolía-P/2019/035 (for JM and QL), and by the Defense Advanced Research Projects Agency (DARPA) under award FA8750-17-C-0106 (for GVS).

\section*{\blue{Appendix A: parameters of the networks and illustrative responses}}
\label{Model_parameters}

\blue{In this Appendix we present the range of parameters that we considered in Eqs.~\ref{Norecur} and~\ref{Recur} of early vision \emph{Model I} and \emph{Model II}. We also show illustrative responses for a natural image}. 

First, note that the throughout the work we consider that the input to our networks is an achromatic image patch of $8\times8$ pixels. This means that vectors $\mathbf{s}$, $\mathbf{x}$, $\mathbf{y}$, $\mathbf{e}$, and $\mathbf{z}$ live in $\mathbb{R}^{64}$, and we consider layers (or nodes) with $n=64$ neurons.
Therefore, matrices $K$, $F$, $\lambda_{CSF}$, and $H$ (that represent relations between neurons) are $64\times64$ matrices.

\begin{figure}[!b]
    \centering
    \includegraphics[width=\textwidth]{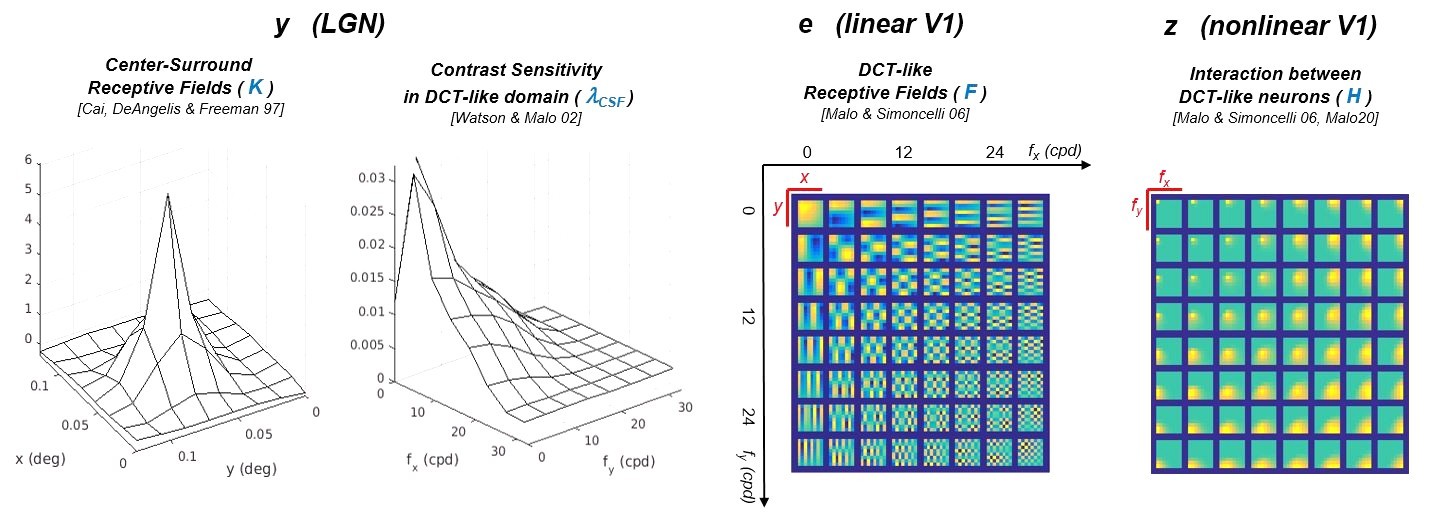}
    \caption{\small{\textbf{Parameters of the vision models:} Center-surround receptive fields in LGN and equivalent Contrast Sensitivity Function. Local frequency filters tuned to different orientations in linear V1 and interaction kernel $H_{ff'}$ in the divisive normalization nonlinearity in V1.}}
    \label{fig:parameters}
\end{figure}

Figure~\ref{fig:parameters} illustrates 
the parameters involved in the retina-to-LGN transform ($\mathbf{x}\rightarrow\mathbf{y}$) and in the LGN-to-cortex transform ($\mathbf{y}\rightarrow\mathbf{z}$), as well as in the intra-cortical nonlinearity ($\mathbf{e}\rightarrow\mathbf{z}$) of \emph{Model I}.

First, regarding $\mathbf{x}\rightarrow\mathbf{y}$ we  follow the relation between the center-surround cells in LGN and the CSF, and hence we compute $K$ from the CSF of the Standard Spatial Observer~\cite{Watson02} transformed from the original Fourier domain into the (more convenient) DCT domain using the procedure in~\cite{Malo97} (second panel in Fig.~\ref{fig:parameters}).
The result (in the spatial domain) are center-surround receptive fields which are consistent with the physiological measurements~\cite{DeAngelis97} (first panel in~\ref{fig:parameters}).

Then, the linear cortical transform 
$\mathbf{y}\rightarrow\mathbf{e}$
uses the local-DCT representation following previous results on biologically-inspired image compression~\cite{Watson94,Ahumada97}
and subjective image quality~\cite{Watson93,Malo97subjective}. 
The $64\times64$ local-frequency receptive fields in $F$ (DCT-like basis functions) are shown in the third panel of Fig.~\ref{fig:parameters}.

Finally, regarding the intra-cortical Divisive Normalization, $\mathbf{e}\rightarrow\mathbf{z}$, here we also follow models used in biologically-inspired
image compression methods~\cite{malo06a,camps-valls2008a}. In this case, the structural connectivity between different local-frequency sensors decays with distance in frequency according to a Gaussian~\cite{Watson97,Martinez18}:
\begin{equation}
    H_{ff'} = e^{-\frac{(f-f')^2}{\sigma(f)^2}}
\end{equation}
where the width $\sigma(f)$ increases with the frequency $f$, according to $\sigma(f) = \sigma_0 + \alpha_H f$, as illustrated in the example of the fourth panel of Fig.~\ref{fig:parameters}. In that case, the connectivity neighborhood is wider for sensors of high frequency (bottom right of the plot) than for sensors of low frequency (top left of the plot).
Finally, in our experiments we set the semi-saturation constant $b$ and the constant $\kappa$ according to the method in~\cite{Gomez19} so that the Divisive Normalization is compatible with classical non-linearities such as the Wilson-Cowan recurrent model~\cite{Wilson73}.

In the experiments we consider a range of intra-cortical connectivity values in \emph{Model I} (section~\ref{res_anal_1}), and we 
modify the width of the kernel $H$ by varying the constant $\alpha_H \in [0.35,4]$, and by varying the strength $c_{ez} \in [0.01,300]$. This has an effect in the nonlinearity of the cortical responses and, as a consequence, on the statistical effect of $f(\cdot)$.   

\begin{figure}[!ht]
    \centering
    \includegraphics[width=\textwidth]{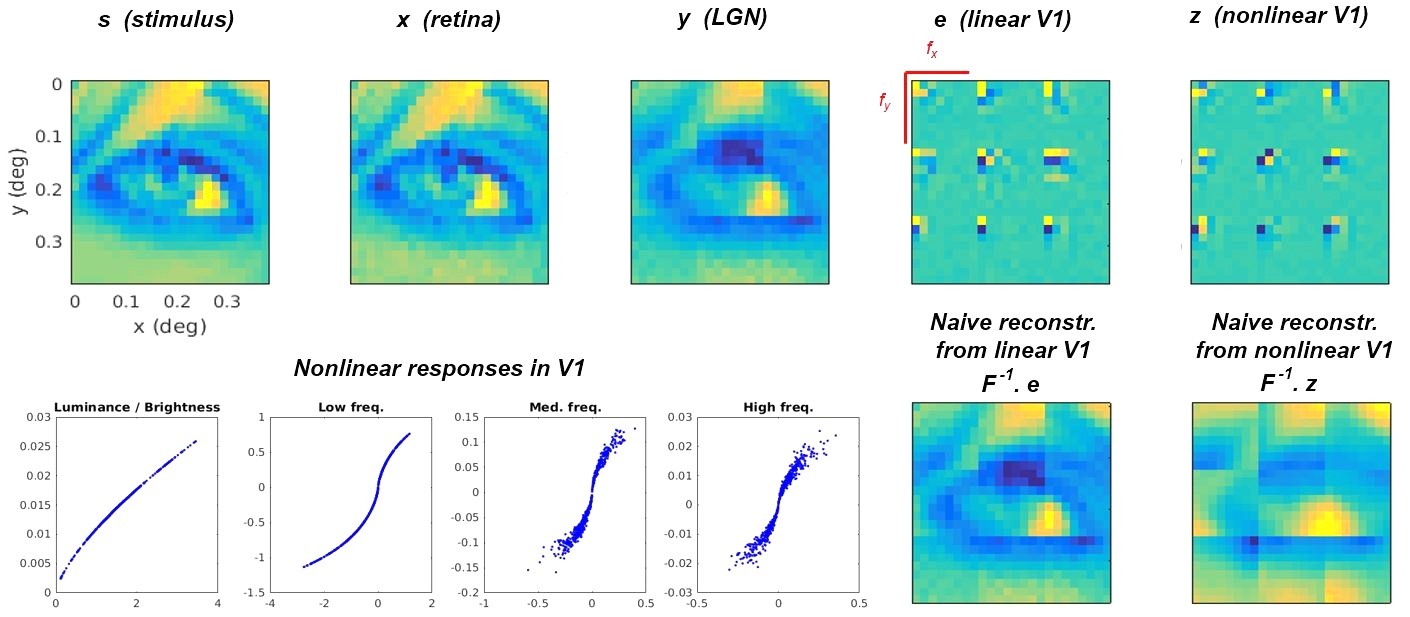}
    \caption{\small{\textbf{Signals through the layers of the vision \emph{Model I}}. Responses to a sample image with the optimal  parameters. Optimal means maximum correlation with human opinion  among the considered discrete set of connectivity values (see Appendix B).}}
    \label{fig:responses}
\end{figure}

Figure~\ref{fig:responses} illustrates the transformations of the signal along the layers of \emph{Model I} for a representative set of parameters (those that maximize correlation with human psychophysics). The top panel shows 
(i) the input image $\mathbf{s}$: in this case the achromatic image of an eye in the range [0,200] $cd/m^2$, spatially sampled at 64~cycles/degree, 
(ii) how this input is distorted with the noise at the retina (leading to $\mathbf{x}$), 
(iii) the response of center-surround cells distorted by noise in $\mathbf{y}$, 
(iv) the response to $3\times3$ regions of local-frequency sensors in $\mathbf{e}$ (with the corresponding noise) in $\mathbf{e}$, and finally,
(v) the result of the Divisive Normalization in $\mathbf{z}$.
Additionally, for a qualitative understanding of the information lost along the way,  the cortical signals ($\mathbf{e}$ and $\mathbf{z}$) are represented back in the spatial domain by transforming them using the linear inverse $F^{-1}$.

Following the argument in~\cite{Esteve20}
the standard deviation of the noise injected at each layer has been selected such as it remains barely visible. This is because just-noticeable-differences are determined by this amount of noise~\cite{Messe06}. Specifically, the standard deviation of the white noise at the different layers in \emph{Model I} is $\sigma(n_x)= 5 cd/m^2$ (for images with luminance in the range $[0, 200] cd/m^2$), $\sigma(n_y)=0.1$, $\sigma(n_e)=0.01$, and (on top of these values), in \emph{Model II} we have $\sigma(n_z)=0.01$.

Finally, the scatter plots at the bottom left of Fig.~\ref{fig:responses} illustrate the nonlinearities introduced by the considered Divisive Normalization.
From the local DC components of the representation we can see the saturation of (perceived) brightness as a function of the input luminance, where we can see the Weber Law~\cite{Fairchild13}.
Similarly, the other plots for \emph{low}, \emph{medium}, and \emph{high}, frequency coefficients, illustrate the nonlinearity of the perceived contrast as a function of the input contrasts. This sigmoidal and signal-dependent behavior is consistent with the psychophysics of contrast perception~\cite{Watson97}, and the amplitide of the responses for the different frequencies is consistent with the CSF~\cite{Li22csfs,Malo97}. 

\section*{\blue{Appendix B: psychophysical plausibility of the networks}}
\label{Models_plausibility}

\blue{In this Appendix we assess the plausibility of the models according to their ability to predict experimental psychophysical data on subjective image quality.}

Note that qualitative Weber law, saturation of perceived contrast, and compatibility with the CSF displayed in Fig.~\ref{fig:responses} suggest that the parameters we chose make biological sense. 
However, a more comprehensive/quantitative test is necessary particularly if a range of parameters has to be considered. 
To this end, here we use the networks with different parameters to predict subjective image quality data, specifically the ratings given by humans in the TID database~\cite{Ponomarenko08}. 
This way of determining plausible parameters is not new~\cite{Watson02,Laparra10,Malo10} and it has been subject to criticism as a single measurement of performance~\cite{Martinez19}. However, in the context presented here, prediction of subjective quality is enough to highlight the general behavior of the models and to (roughly) identify which regions of the parameter space make more biological sense.

\begin{figure}[!ht]
    \centering
    \includegraphics[width=0.9\textwidth]{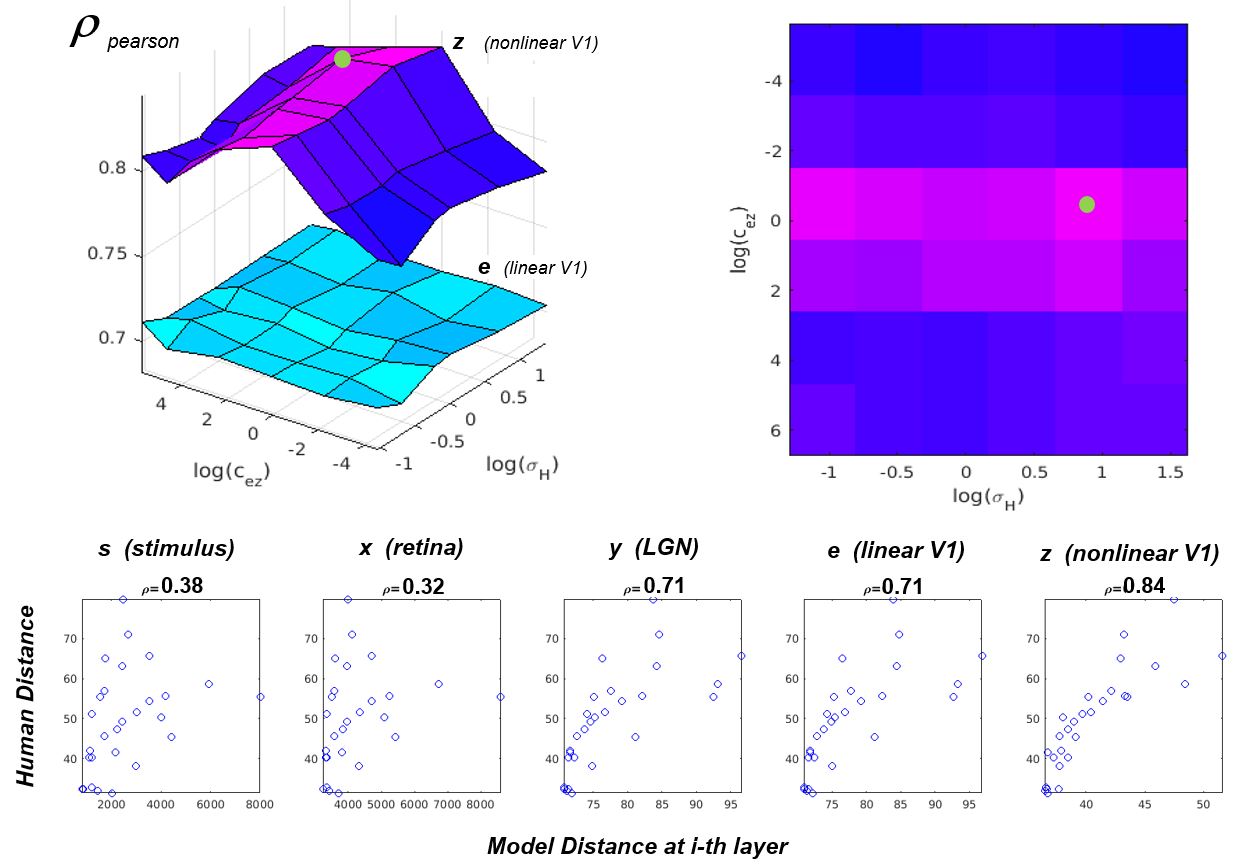}
    \caption{\small{\textbf{Vision Models \emph{I} and \emph{II} are psychophysically plausible.} Correlation with human opinion for different cortical connectivity values (surfaces on top) and correlations in previous (linear) layers (scatter plots at the bottom). In the nonlinear cortical case the scatter plot is the one corresponding to the optimum connectivity.}}
    \label{fig:plausibility}
\end{figure}

In this regard, the scatter plots in Figure~\ref{fig:plausibility} show how well Euclidean distances at the different layers of \emph{Model I} (abscisas), predict the subjective ratings (ordinates).
The strong correlation obtained in the inner cortical representation $\rho=0.84$, which is not far from the state-of-the-art in subjective image quality metrics~\cite{Hepburn2020} prove the plausibility of the transforms and the levels of the Gaussian noise introduced at each layer.

Specifically, the poor result for the input representation ($\mathbf{s}$ in luminance) implies that the visual brain certainly \emph{does something} to the input signal~\cite{Teo94,Wang09}. 
The progressive improvement of the correlation along deeper layers means that the set of considered transforms is biologically meaningful. In fact, the consideration of the center-surround cells (or the CSF) is a major fact in explaining image quality~\cite{Watson93,Watson02}, and this is incorporated in both \emph{Model I} and \emph{Model II} leading to a reasonable Pearson correlation, $\rho = 0.71$, only with linear transforms. Then, we study the intra-cortical connectivity of \emph{Model I} in more detail: we consider the plausibility of a range of strengths $c_{ez}$ and a range of widths in $H$.  

The result shows that all the family of Divisive Normalization transforms make sense because they substantially improve the correlation with human opinion. Note that the correlation at the linear cortical layer $\mathbf{e}$ (surface in light blue at 0.71) is raised by the different $\mathbf{z}$ layers to be in the range $[0.76, 0.84]$. Moreover, the final correlation surface for the different intra-cortical connectivity values has strong curvature and a clear maximum (green dot) in the middle of the considered region. This means that it is interesting to study the behavior of the statistical descriptors of connectivity in this region of parameters.

\section*{\blue{Appendix C: classical correlation as special case of $I$ and $T$}}
\label{Relation_between_concepts}
\blue{The links and differences between the three considered magnitudes (the classical correlation $\rho$, as a measure of the off-diagonal coefficients in the covariance matrix, $I$ and $T$) can be seen from the interesting expression proposed by Cardoso in~\cite{Cardoso03}, where given a random vector, $\vect{a} \in \mathbb{R}^n$, it holds:}
\begin{equation}
       \blue{T(\vect{a}) = C(\vect{a}) + J(\vect{a}) - J_m(\vect{a})}
       \label{cardoso}
\end{equation}
\blue{where, $C(\vect{a})$ summarizes the 2nd-order \emph{linear correlation} by describing the amplitude of the off-diagonal elements of the covariance matrix compared to the diagonal elements:}
\begin{equation}
       \blue{C(\vect{a}) = \frac{1}{2} \, log \, | 2\pi e \,\, \textrm{diag}( \Sigma^{\vect{a}} ) |- \frac{1}{2} \, log \, | 2\pi e \,\,  \Sigma^{\vect{a}} |}
       \nonumber
\end{equation}
\blue{and $J(\vect{a})$ and $J_m(\vect{a})$ are the joint and marginal \emph{negentropies} and they measure the Kullback-Leibler divergences from Gaussian distributions:}
\blue{
\begin{eqnarray}
       J(\vect{a}) &=& D_{\textrm{KL}}(p(\vect{a})|p^G(\vect{a})) \nonumber \\ 
       J_m(\vect{a}) &=& \sum_{i=1}^n D_{\textrm{KL}} (p(a_i)|p^G(a_i)) \nonumber
\end{eqnarray}}
\blue{where $p(\cdot)$ is the PDF (joint or marginal) of the considered (vector or scalar) variable, and $p^G(\cdot)$ is its best Gaussian approximation.}

\blue{Eq.~\ref{cardoso} means that for general (non-Gaussian) PDFs the measure $T(\vect{a})$ includes factors beyond the all the \emph{2nd-order pair-wise linear correlations} represented by $C(\vect{a})$.} 

\blue{The $I$ and $T$ are obviously different because of their \emph{pairwise} versus \emph{multi-way} nature which is critical in this work (Eqs.~\ref{defT}-\ref{defI}). 
Interestingly, this difference still remains even if we consider just two Gaussian variables, and in that situation it is very clear that $T$
captures intra-variable redundancy as opposed to $I$. Moreover, the link to classical correlation is also apparent.}  
%This can be seen using Eq.~\ref{cardoso} for $T$ and the definition of $I$ using the entropy of Gaussian variables. 

\blue{On the one hand, for $T$, if we consider two Gaussian signals, $\vect{a} = [\vect{x} \,\, \vect{y}]$, the negentropies in Eq.~\ref{cardoso} vanish and then:}
\blue{
\begin{equation}
       T(\vect{x},\vect{y}) = C(\vect{x}) + C(\vect{y}) -  \frac{1}{2} log \, \left| \, \mathbb{I} - (\Sigma^{\vect{x}} \cdot \Sigma^{\vect{y}})^{-1} \cdot \mathbb{E}\{\vect{x}\cdot\vect{y}^\top\} \cdot \mathbb{E}\{\vect{y}\cdot\vect{x}^\top \}  \, \right|
       \label{defTg}
\end{equation}
On the other hand, for $I$, using the entropy of the Gaussian variables one has:
\begin{equation}
       I(\vect{x},\vect{y}) = -  \frac{1}{2} log \, \left| \, \mathbb{I} - (\Sigma^{\vect{x}} \cdot \Sigma^{\vect{y}})^{-1} \cdot \mathbb{E}\{\vect{x}\cdot\vect{y}^\top \} \cdot \mathbb{E}\{\vect{y}\cdot\vect{x}^\top \}  \, \right|
       \label{defIg}
\end{equation}}
\blue{which, in case of comparing univariate variables $x$ and $y$, reduces to the standard \emph{linear correlation}:
\begin{equation}
       I(x,y) = -  \frac{1}{2} log \left( 1 - \frac{\sigma_{xy}^2}{\sigma_x^2\sigma_y^2} \right) = -  \frac{1}{2} log \, ( 1 - \rho^2 ) \nonumber
\end{equation}}
\blue{These equations~\ref{defTg}-\ref{defIg} illustrate the different amount of information captured by $T$ and $I$: while $I$ describes the redundancy between the two variables, $T$ also includes the redundancy within each variable. Note that $I$ vanishes if the off-diagonal blocks $\mathbb{E}\{ \vect{x}\cdot \vect{y}^\top \}$ (relations between variables) are zero. However, $T$, on top of this, it also contains $C(\vect{x})$ and $C(\vect{y}$), that describe the relations within the dimensions of $\vect{x}$, and the relations within the dimensions of~$\vect{y}$.
The consideration of intra-variable redundancy in $T$ as opposed to $I$ is not restricted to Gasussian variables (as Eqs.~\ref{defTg}-\ref{defIg}), it is general and can be seen in Venn diagrams of $T$ and $I$~\cite{Laparra20}.}

\section*{\blue{Appendix D: Variability of information estimates}}
\label{significance}

As stated in Sections~\ref{Methods} and~\ref{Empiric_results_real}, the empirical estimations of information in this work are based on the Rotation-Based Iterative Gaussianization (RBIG)~\cite{Laparra11,Laparra20}. We used the option in which rotations are given by the \emph{deterministic} sample covariance at each iteration. With that deterministic choice, given a dataset, the information estimate is unique. However, due to the variability of the data, different sets drawn from the same distribution lead to a distribution (or histogram) of information estimates.

\blue{In this setting, the statistical significance of differences in two information values estimated for two connectivity situations depends on the eventual overlap between the two histograms of information estimates.
This overlap depends on the intrinsic variability of the data and on the sensitivity of the estimate on the finite size of the dataset.}

\begin{figure}[h]
\vspace{-0.25cm}
    \centering
    \includegraphics[width=0.78\textwidth, height=12cm]{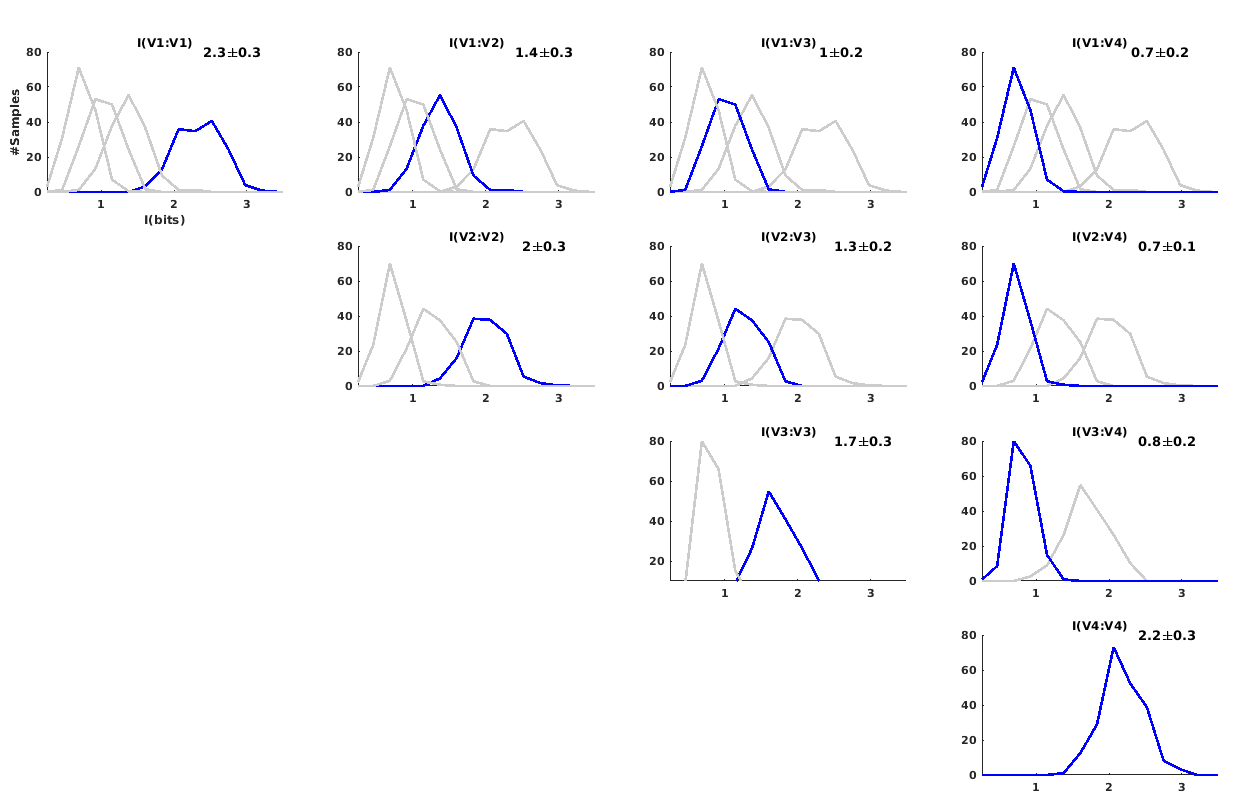}\\
     \begin{center}
       \line(1,0){370}
     \end{center}    
    \includegraphics[width=0.78\textwidth, height=3cm]{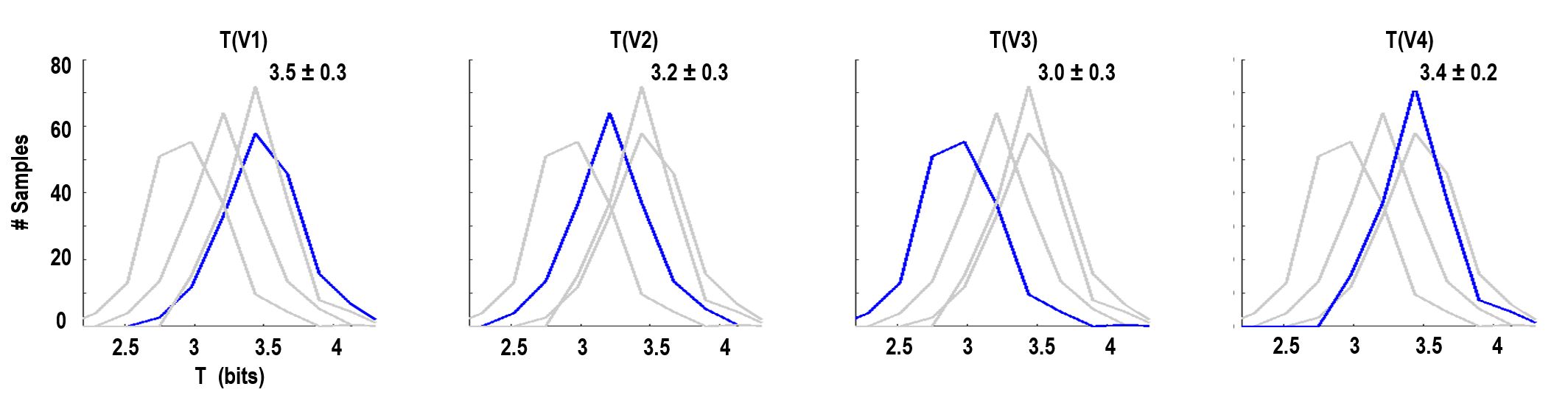}
    \caption{\small{\textbf{Information distributions corresponding to the fMRI data (first and second panels of Table 1)}} top and bottom respectively. In each plot the histogram \emph{in blue} corresponds to the nodes that are being considered in the corresponding cell of the table. The histograms \emph{in gray} are those in the same row of the table and are just drawn as reference to visualize the variation of the estimations in blue.}
    \label{histog1}
\end{figure}

\begin{figure}[h]
\vspace{-0.25cm}
    \centering
    \includegraphics[width=0.78\textwidth, height=10cm]{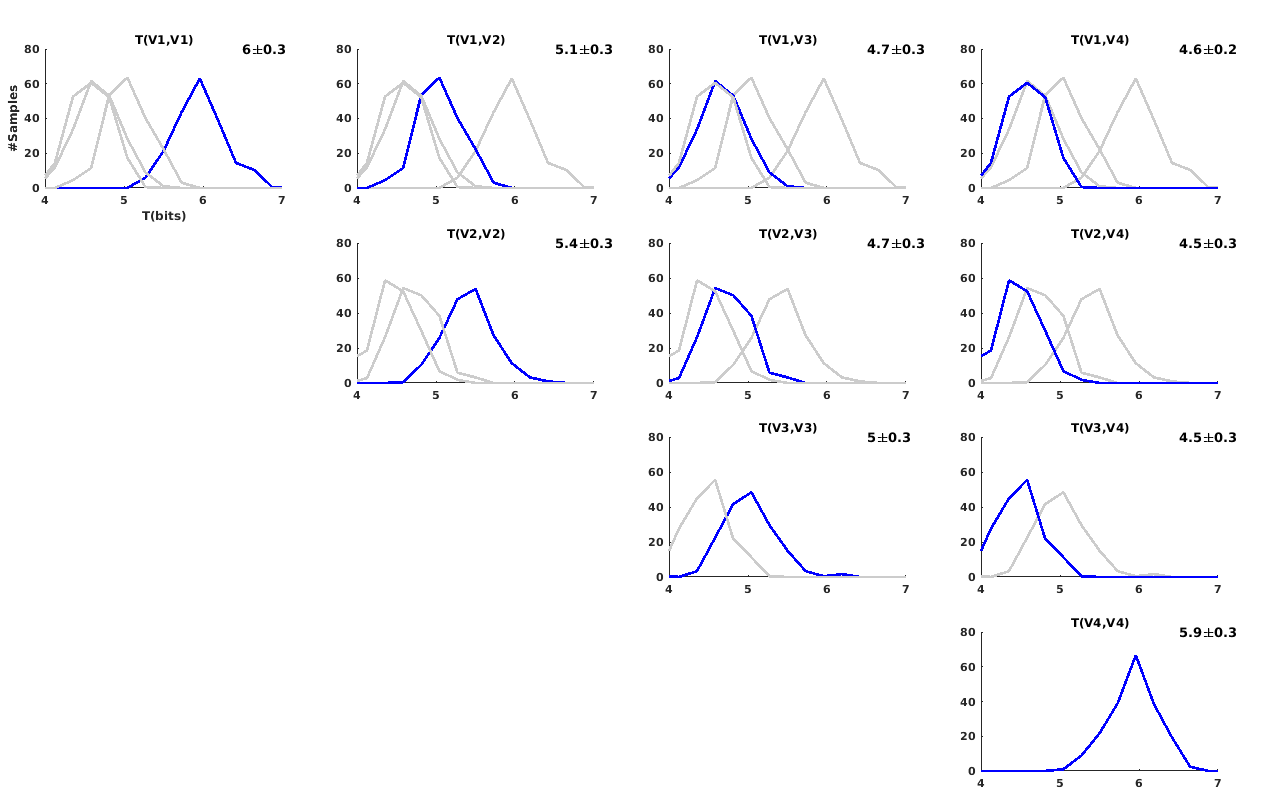}\\\vspace{-0.5cm}
    %\hline\\\hline\\
     \begin{center}
       \line(1,0){370}
     \end{center}    
    \includegraphics[width=0.78\textwidth, height=5.5cm]{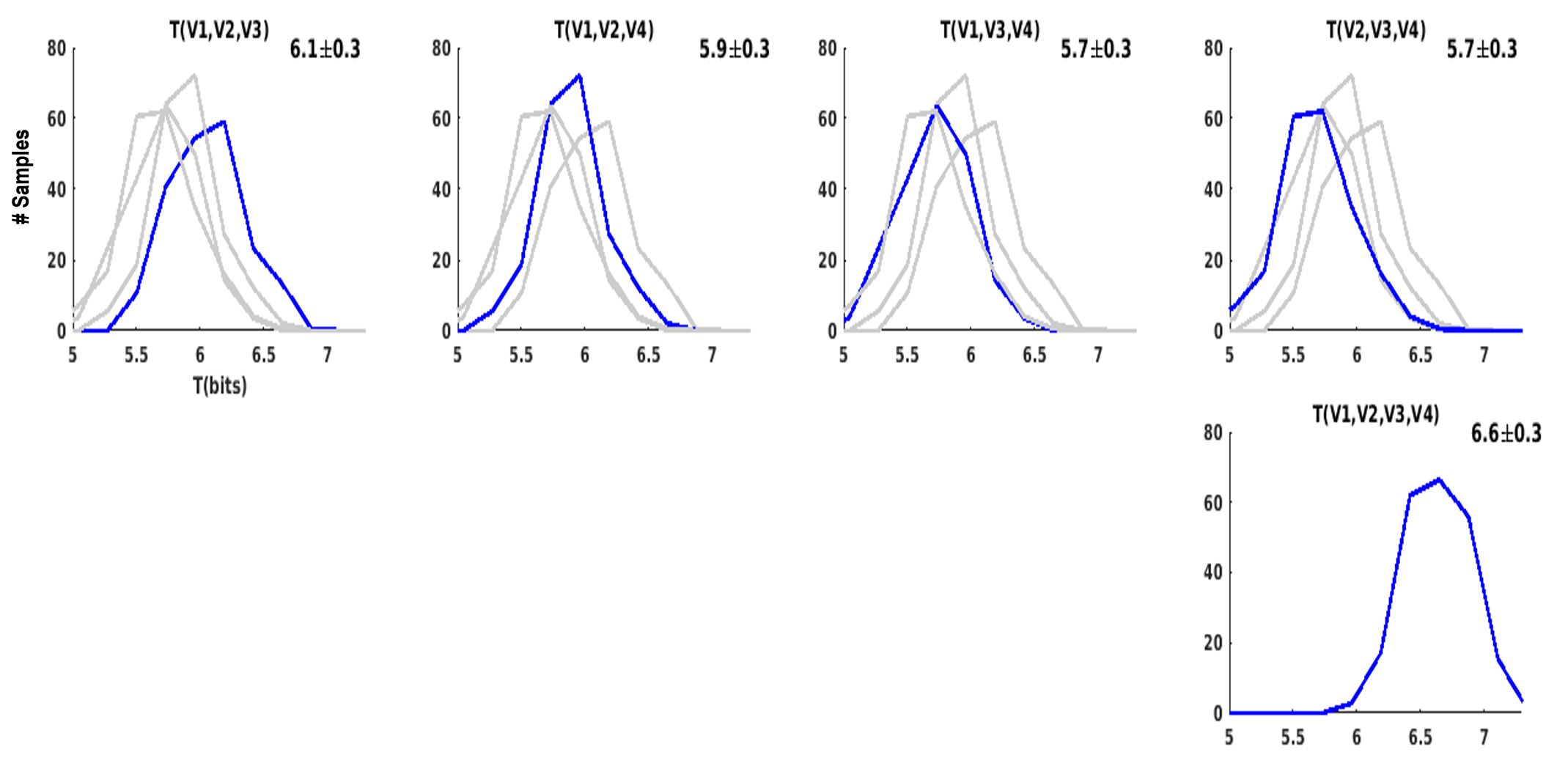} \caption{\small{\textbf{Information distributions corresponding to the fMRI data (third panel of Table 1)}. The array on the top corresponds to the pair-wise cases and below the line we have the cases with 3 and 4 nodes. In each plot the histogram \emph{in blue} corresponds to the nodes that are being considered. The histograms \emph{in gray} are those with the same number of nodes. They are just drawn as reference to visualize the variation of the estimations in blue.}}
    \label{histog2}
\end{figure}

\begin{figure}[ht!]
\vspace{-0.25cm}
    \centering
    \includegraphics[width=0.45\textwidth, height=6.5cm]{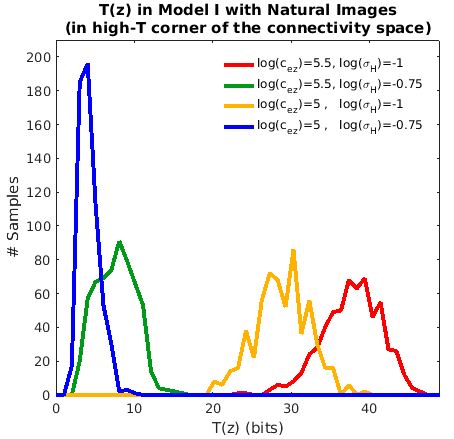} \caption{\blue{\small{\textbf{Illustrative information distributions corresponding to four points of the high-T corner of the surface in Fig.~\ref{fig:T_forward}}. The location of the four points in the space of connectivity parameters is given in the legend.}}}
    \label{hist_natur}
\end{figure}

\blue{In this appendix we illustrate the variability of the RBIG information 
estimates by showing:}
\begin{itemize}

\item \blue{All the distributions for the information estimations for the case of fMRI responses from visual regions V1-V4  (Section 5.4) in Figs.~\ref{histog1}-\ref{histog2}.}

\item \blue{The distributions obtained for four representative estimations of $T$ in the case of natural image data through the analytical \emph{Model I} (Section 5.2) in Fig.~\ref{hist_natur}.}
\end{itemize}

For the fMRI data the variability comes from random selection of the voxels from the different regions. This selection is independent in each realization and, in this real scenario there is no guarantee that the selected voxels have relevant connections.
On the other hand, in the analytical model the correspondence is known. Therefore all the variability comes from the different behavior of sets of natural images in our synthetic V1-cortex.

For the fMRI data, the consistent shifts of the histograms of RBIG estimations confirm the statements about the reduction of transmitted information along the layers (from V1 to V4), the redundancy reduction within layers (from V1 to V3), and the bigger connectivity among V1-V3 than among V2-V4.  Conventional tests based on the overlapping of the histograms could quantify the significance of the differences, but this is already noticeable from the standard deviations in the tables. In particular, the reduction of information shared between progressively distant nodes is particularly significant, while the difference between the connectivity in V1-V3 and V2-V4 is less significant due to the uncertainty introduced by the random selection of voxels. 

On the other hand, the moderate overlap of histograms in Fig.~\ref{hist_natur} shows that the results for the analytical network  in Fig.~\ref{fig:T_forward}, are actually different. Moreover, the standard error of the means (with 500 realizations), would be very small.

\section*{\blaveros{Appendix E: illustrative results with alternative descriptors}}
\label{alternative}

\blaveros{As a convenient reference, in this Appendix we illustrate the ability of other descriptors of connectivity (partial cross correlation~\cite{PartialCrossCorrelation05}, Coherence~\cite{Sun04}, and phase locking value~\cite{PhaseLock99}) to capture the variations of connectivity in one of the considered experiments. 
Theoretical considerations about these methods are out of the scope of this work, so in this appendix we take a purely empirical approach: we just apply the default options of a recent implementation of these measures~\cite{Zhou09} for the nodes $\vect{x}$ and $\vect{z}$ in the range of connectivity values of the nonlinearity of Model I (a subset of the experiments in section 4.2), both for Gaussian signals and Natural Images.} 

\begin{figure}[ht!]
\vspace{-0.25cm}
    \centering
    \includegraphics[width=0.75\textwidth, height=6cm]{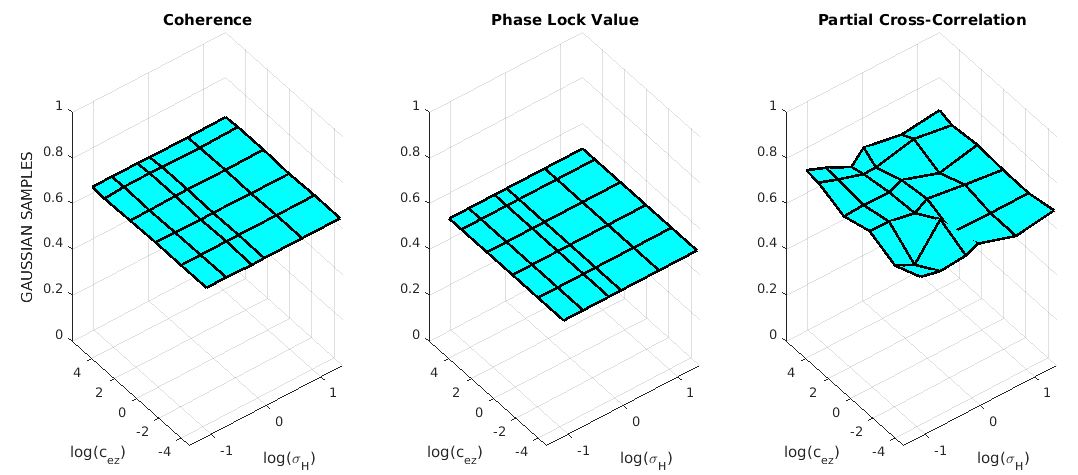}\\
    \includegraphics[width=0.75\textwidth, height=6cm]{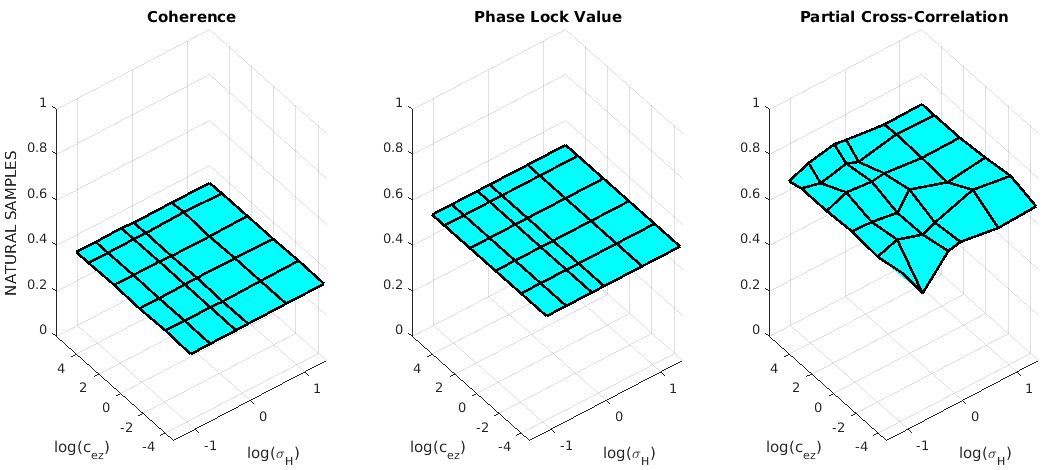}\\    
    \caption{\blaveros{\small{\textbf{Alternative measures do not capture variations in connectivity in Divisive Normalization (Model I)}. Connectivity measures between nodes $\vect{x}$ and $\vect{z}$  for a range of interactions ($c_{ez}$, $\sigma_H$) in the nonlinearity for Gaussian Signals (top) and Natural Images (bottom).}}}
    \label{alterna}
\end{figure}

\blaveros{Fig.~\ref{alterna} shows that none of these metrics captures the variability of connectivity along the considered nonlinear models.
This inability is similar to the poor results of Mutual Information
(in Fig. 3). A specific theoretical analysis of each method would be required to explain this inability (as we did for $I$). This is out of the scope of this work, but this empirical exploration of alternative descriptors stresses the interest of the sensitivity of $T$ shown in Eqs. 8-10, and in Fig. 4.} 

\clearpage
%\newpage
%\bibliographystyle{unsrt}
%\bibliography{references}
\medskip
\printbibliography

@article{Gomez19,
  title={Visual Information Flow in {Wilson-Cowan} Networks},
  author={Gomez-Villa, A. and Bertalmio, M. and Malo, J.},
  journal={J. Neurophysiol. doi:10.1152/jn.00487.2019},
  year={2020}
}

@article{PartialCrossCorrelation05,
title = {Undirected graphs of frequency-dependent functional connectivity in whole brain networks},
journal = {Phil. Trans. Roy. Soc. B: Biol. Sci.},
volume = {360},
number = {1457},
pages = {937–946},
year = {2005},
author = {Salvador, R. and Suckling, J. and Schwarzbauer, C. and  Bullmore, E.},
}

@article{PhaseLock99,
author = {Lachaux, JP. and Rodriguez, E. and Martinerie, J. and Varela, FJ.},
title = {Measuring phase synchrony in brain signals},
journal = {Human Brain Mapping},
volume = {8},
number = {4},
pages = {194-208},
year = {1999}
}

@article{Ivanov69,
  title={Method of expressing the sensitivity of measuring and recording apparatus},
  author={Ivanov, M.P.},
  journal={ Measur. Technics},
  year={1969},
  volume={12},
  pages={762–764}
}

@article{Mandel54,
  title={Sensitivity--a criterion for the comparison of methods of test},
  author={Mandel, J. and Stiehler, R.D.},
  journal={J. Res. Nat. Bur. Stand.},
  year={1954},
  volume={53},
  pages={155-159}
}

@article{Zhou09,
  title={{MATLAB} toolbox for functional connectivity},
  author={Zhou, D. and Thompson, W. K. and Siegle, G.},
  journal={NeuroImage},
  year={2009},
  volume={47},
  number={4},
  pages={1590–1607}
}

@article{Mohanty20,
title = {Rethinking Measures of Functional Connectivity via Feature Extraction},
journal = {Sci. Rep.},
author = {Mohanty, R. and Sethares, W.A. and Nair, V.A. and others},
volume = {10},
pages = {1298},
year = {2020}
}

@article{Schreiber00,
  title = {Measuring Information Transfer},
  author = {Schreiber, Thomas},
  journal = {Phys. Rev. Lett.},
  volume = {85},
  issue = {2},
  pages = {461--464},
  year = {2000}
}

@Inbook{Gastpar13,
author="Gastpar, Michael C.",
editor="Jaeger, D.
and Jung, R.",
title="Directed Information Flow and Causality in Neural Systems",
bookTitle="Encycl. Comput. Neurosci.",
year="2013",
publisher="Springer",
address="New York, NY",
pages="1--3",
}

@Inbook{Massey90,
author="Massey, JL.",
title="Causality, feedback and directed information",
bookTitle="Proc. of the 1990 Intl. symp. Inf. Theory and Appl.",
year="1990",
address="Hawaii",
pages="303--305",
}

@Article{Tashi23,
AUTHOR = {Namgyal, T. and Hepburn, A. and Santos-Rodriguez, R. and Laparra, V. and Malo, J.},
TITLE = {What You Hear Is What You See: Audio Quality Metrics From Image Quality Metrics},
JOURNAL = {26th Int. Conf. Digital Audio Effects. arxiv 2305.11582},
YEAR = {2023},
}

@Article{Malo22,
AUTHOR = {Malo, Jesús},
TITLE = {Information Flow in Biological Networks for Color Vision},
JOURNAL = {Entropy},
VOLUME = {24},
YEAR = {2022},
NUMBER = {10},
ARTICLE-NUMBER = {1442},
DOI = {10.3390/e24101442}
}

@article{AudioAdapt,
  title={Adaptation in auditory processing},
  author={Willmore, B. and King, A.},
  journal={Physiol.Rev.},
  volume={103},
  number={2},
  pages={1025--1058},
  year={2023}
}

@article{Ma08,
title = {Overcomplete topographic independent component analysis},
journal = {Neurocomputing},
volume = {71},
number = {10},
pages = {2217-2223},
year = {2008},
author = {Ma, L. and Zhang, L.},
}

@article{Hernandez23,
title = {Neural networks with divisive normalization for image segmentation},
journal = {Pattern Recognition Letters},
volume = {173},
pages = {64-71},
year = {2023},
author = {Hernández-Cámara, P. and Vila-Tomás, J. and Laparra, V. and Malo, J.},
}

@article{Hyvarinen01,
title = {Topographic independent component analysis as a model of {V1} organization and receptive fields},
journal = {Neurocomputing},
volume = {38-40},
pages = {1307-1315},
year = {2001},
author = {Hyvärinen, A. and Hoyer, P.O.},
}

@inproceedings{alexnet,
author = {Krizhevsky, A. and Sutskever, I. and Hinton, G.E.},
title = {ImageNet Classification with Deep Convolutional Neural Networks},
year = {2012},
booktitle = {Proc. 25th Neural Inf. Proc. Syst.},
pages = {1097–1105}}

@InProceedings{vgg16,
author={Simonyan, K. and Zisserman, A.},
title={Very Deep Convolutional Networks for Large-Scale Image Recognition},
booktitle= {Proc. 3rd Int. Conf. Learn. Repr.},
year={2015},
pages = {1-14}}

@article{Barnett09,
  title = {Granger Causality and Transfer Entropy Are Equivalent for Gaussian Variables},
  author = {Barnett, L. and Barrett, A.B. and Seth, A.K.},
  journal = {Phys. Rev. Lett.},
  volume = {103},
  issue = {23},
  pages = {238701},
  numpages = {4},
  year = {2009},
  month = {Dec}
}

@article{Battiston20,
title = {Networks beyond pairwise interactions: Structure and dynamics},
journal = {Physics Reports},
volume = {874},
pages = {1-92},
year = {2020},
note = {Networks beyond pairwise interactions: Structure and dynamics},
author = {Battiston, F. and others},
}

@article{Sun04,
title = {Measuring interregional functional connectivity using coherence and partial coherence analyses of fMRI data},
journal = {NeuroImage},
volume = {21},
number = {2},
pages = {647-658},
year = {2004},
author = {Felice T Sun and Lee M Miller and Mark D'Esposito}
}

@book{Tekalp15,
  title={Digital Video Processing},
  author={Tekalp, A.M.},
  series={Prentice Hall Signal Processing Series},
  year={2015},
  publisher={Prentice Hall}
}

@article{Herzog22,
title = {Genuine high-order interactions in brain networks and neurodegeneration},
journal = {Neurobiology of Disease},
volume = {175},
pages = {105918},
year = {2022},
author = {Herzog, R. and Rosas, FE. and Whelan, R. and  Fittipaldi, S. and Santamaria-Garcia, H. and Cruzat, J. and Birba, A. and Moguilner, S. and Tagliazucchi, E. and Prado, P. and Ibanez, A.},
}

@article{Gatica21,
title = {High-order Interdependencies in the aging brain},
journal = {Brain Connect.},
volume = {11},
number = {9},
pages = {734-744},
year = {2021},
author = {Gatica, M. and Cofre, R. and Mediano, PAM. and  Rosas, FE. and Orio, P. and Diez, I. and  Swinnen, SP. and Cortes JM.},
}

@article{NaturalScenes_21,
  title={A massive 7{T} f{MRI} dataset to bridge cognitive neuroscience and artificial intelligence.},
  author={Allen, E.J. and St-Yves, G. and Wu, Y.},
  journal={Nature Neurosci.},
	year = {2022},
    volume = {25},
	pages = {116–126}
}

@article{BOLD5000_19,
  title={BOLD5000, a public fMRI dataset while viewing 5000 visual images.},
  author={Chang, N. and Pyles, J.A. and Marcus, A.},
  journal={Sci. Data},
	year = {2019},
    volume = {6},
	pages = {49}
}

@ARTICLE{Marin-Franch13,
  author={Marín-Franch, Iván and Foster, David H.},
  journal={IEEE Trans. Patt. Anal. Mach. Intell.}, 
  title={Estimating Information from Image Colors: An Application to Digital Cameras and Natural Scenes}, 
  year={2013},
  volume={35},
  number={1},
  pages={78-91}
}

@article{Li22csfs,
    author = {Li, Qiang and Gomez-Villa, Alex and Bertalmío, Marcelo and Malo, Jesús},
    title = "{Contrast sensitivity functions in autoencoders}",
    journal = {Journal of Vision},
    volume = {22},
    number = {6},
    pages = {8-8},
    year = {2022},
    month = {05}
}

@article{Hancock92,
author = {Hancock, P.J.B. and Baddeley, R.J. and Smith, L.S. },
title = { The principal components of natural images},
journal = {Network},
volume = {3},
pages = {61–70},
year = {1992},
}

@article{Laparra15,
  title={Visual aftereffects and sensory nonlinearities from a single statistical framework},
  author={Laparra, V. and Malo, J.},
JOURNAL={Front. Human Neurosci.},
VOLUME={9},
PAGES={557},
YEAR={2015},
}

@article{Atick92,
author = {Atick, J. and Li, Z. and Redlich, A.},
title = {Understanding Retinal Color Coding from First Principles},
journal = {Neural Comput.},
volume = {4},
number = {4},
pages = {559-572},
year = {1992},
}

@ARTICLE{Barlow01,
	author = {Barlow, H.},
	title = {Redundancy reduction revisited},
	journal = {Network: Comp. Neur. Syst.},
	year = {2001},
    volume = {12},
	number = {3},
	pages = {241-253}
}

@article{Simoncelli98,
author = "Simoncelli, E. and Heeger, D.",
title = "A model of neuronal responses in visual area {MT}",
journal = "Vis. Res.",
volume = "38",
number = "5",
pages = "743 - 761",
year = "1998",
}

@ARTICLE{Brainard05,
  AUTHOR =       {J. M. Hillis and D.H Brainard},
  TITLE =        {Do common mechanisms of adaptation mediate color discrimination and appearance?},
  JOURNAL =      {JOSA A},
  volume =       {22},
  number =       {10},
  pages =        {2090--2106},
  year =         {2005},
}

@article{Parga10,
title = {The asynchronous state in cortical circuits},
author = {Renart, A. and Rocha, J. and Bartho, P. and Hollender, L. and Parga, N. and Reyes, A. and Harris, KD.},
journal = {Science},
volume = {327(5965)},
pages = {587-590},
year = {2010},
}

@article{Li22,
title = {Functional connectivity inference from fMRI data using multivariate information measures},
author = {Qiang Li},
journal = {Neural Networks},
volume = {146},
pages = {85-97},
year = {2022},
}

@ARTICLE{Carandini94,
  AUTHOR =       "M. Carandini and D. Heeger",
  TITLE =        "Summation and Division by Neurons in Visual Cortex",
  JOURNAL =      "Science",
  YEAR =         "1994",
  volume =       "264",
  number =       "5163",
  pages =        "1333--6"
}

@ARTICLE{Rust06,
  AUTHOR =       "Rust, N.C., and Movshon, J.A.",
  TITLE =        "In praise of artifice",
  JOURNAL =      "Nat. Neurosci.",
  YEAR =         "2005",
  volume =       "8",
  pages =        "1647–1650"
}

@article{Carandini12,
  title={Normalization as a canonical neural computation},
  author={Carandini, M. and Heeger, D.},
  journal={Nat. Rev. Neurosci.},
  volume={13},
  number={1},
  pages={51-62},
  year={2012}
}

@article{Lizier11,
author = {Lizier, J.T. and Heinzle, J. and Horstmann, A. and Haynes, J. and Prokopenko, M.},
title = {Multivariate Information-Theoretic Measures Reveal Directed Information Structure and Task Relevant Changes in f{MRI} Connectivity},
year = {2011},
volume = {30},
number = {1},
journal = {J. Comput. Neurosci.},
pages = {85–107},
numpages = {23}
}

@article{Olshausen01,
author = {Eero P Simoncelli and  Bruno A Olshausen},
title = {Natural Image Statistics and Neural Representation},
journal = {Annual Review of Neuroscience},
volume = {24},
number = {1},
pages = {1193-1216},
year = {2001},
}

@article{Martinez19,
  author={Martinez, M. and Bertalm\'io, M and Malo, J.},
  title={In Praise of Artifice Reloaded: Caution with Natural Image Databases in Modeling Vision},
  journal={Front. Neurosci. doi: 10.3389/fnins.2019.00008},
  year={2019},
}

@article{Wilson73,
  title={A mathematical theory of the functional dynamics of cortical and thalamic nervous tissue},
  author={Wilson, Hugh R and Cowan, Jack D},
  journal={Kybernetik},
  volume={13},
  number={2},
  pages={55--80},
  year={1973},
  publisher={Springer}
}

@article{Martinez18,
    author = {Martinez, M. AND Cyriac, P. AND Batard, T. AND Bertalm{\'\i}o, M. AND Malo, J.},
    journal = {PLOS ONE},
    title = {Derivatives and inverse of cascaded linear+nonlinear neural models},
    year = {2018},
    month = {10},
    volume = {13},
    pages = {1-49},
    number = {10},
}

@article{QiangEntr22,
AUTHOR = {Li, Qiang and Steeg, Greg Ver and Yu, Shujian and Malo, Jesus},
TITLE = {Functional Connectome of the Human Brain with Total Correlation},
JOURNAL = {Entropy},
VOLUME = {24},
YEAR = {2022},
NUMBER = {12},
ARTICLE-NUMBER = {1725}
}

@article{Lyu09,
author = {Lyu, Siwei and Simoncelli, Eero P.},
journal = {Neural Computation},
number = {6},
pages = {1485--1519},
title = {{Nonlinear Extraction of Independent Components of Natural Images Using Radial Gaussianization}},
volume = {21},
year = {2009}
}

@article{Cardoso03,
  author    = {Jean{-}Fran{\c{c}}ois Cardoso},
  title     = {Dependence, Correlation and Gaussianity in Independent Component Analysis},
  journal   = {J. Mach. Learn. Res.},
  volume    = {4},
  pages     = {1177--1203},
  year      = {2003},
}

@article{Malo20,
  title={Spatio-chromatic information available from different neural layers via Gaussianization},
  author={Malo, J.},
  journal={J. Math. Neurosci.},
  volume = {10},
  number = {18},
  year={2020}
}

@inproceedings{Johnson19,
  author    = {Johnson, J.E. and
               Laparra, V. and
               Santos, R. and Camps, G. and Malo, J.},
  title     = {Information Theory in Density Destructors},
  booktitle = {7th Int. Conf. Mach. Learn., {ICML} 2019, Workshop on Invertible Normalization Flows},
  year      = {2019}
}

@inproceedings{FeiFei09,
 author = {Chai, B. and Walther, D. and Beck, D. and Fei-fei, L.},
 booktitle = {NIPS},
 editor = {Y. Bengio},
 pages = {270--278},
 publisher = {Curran Associates, Inc.},
 title = {Exploring Functional Connectivities of the Human Brain using Multivariate Information Analysis},
  volume = {22},
 year = {2009}
}

@article{Twer01,
author = {T.v.d. Twer and D.I.A. MacLeod},
title = {Optimal nonlinear codes for the perception of natural colours},
journal = {Network: Computation in Neural Systems},
volume = {12},
number = {3},
pages = {395-407},
year  = {2001},
publisher = {Taylor & Francis}
}

@CONFERENCE{FosterCIC08,
  author =       {Foster, D.H. and Marin-Franch, I. and  Nascimento, S.M.C.},
  title =        {Coding efficiency of CIE color spaces},
  booktitle =    {Proc. 16th Color Imag. Conf.},
  year =         {2008},
  pages  =       {285--288},
  organization = {Soc. Imag. Sci. Tech.},
}

@article{Serences14,
	author = {Saproo, Sameer and Serences, John T.},
	title = {Attention Improves Transfer of Motion Information between V1 and MT},
	volume = {34},
	number = {10},
	pages = {3586--3596},
	year = {2014},
	journal = {J. Neurosci.}
}

@article{Watanabe60,
  title={Information theoretical analysis of multivariate correlation},
  author={Watanabe, Satosi},
  journal={IBM Journal of research and development},
  volume={4},
  number={1},
  pages={66--82},
  year={1960},
  publisher={IBM}
}

@article{Laparra11,
  title={Iterative gaussianization: from {ICA} to random rotations},
  author={Laparra, V. and Camps-Valls, G. and Malo, J.},
  journal={IEEE Trans. Neural Networks},
  volume={22},
  number={4},
  pages={537--549},
  year={2011},
  publisher={IEEE}
}

@article{Kraskov04,
  title = {Estimating mutual information},
  author = {Kraskov, Alexander and St\"ogbauer, Harald and Grassberger, Peter},
  journal = {Phys. Rev. E},
  volume = {69},
  issue = {6},
  pages = {066138},
  numpages = {16},
  year = {2004},
  month = {Jun},
  publisher = {American Physical Society},
}

@inproceedings{steeg17,
	Author = {Ver Steeg, Greg},
	Booktitle = {IJCAI},
	Title = {Unsupervised Learning via Total Correlation Explanation},
	Year = {2017}}

@article{Friston11,
author = {Friston, Karl},
year = {2011},
pages = {13-36},
title = {Functional and Effective Connectivity: A Review},
volume = {1},
journal = {Brain Connect.},
}

@INPROCEEDINGS{Ponomarenko08,
  author={Ponomarenko, N. and Lukin, V. and Egiazarian, K. and Astola, J. and Carli, M. and Battisti, F.},
  booktitle={2008 IEEE 10th Workshop on Multimedia Signal Processing}, 
  title={Color image database for evaluation of image quality metrics}, 
  year={2008},
  volume={},
  number={},
  pages={403-408},
  }

@conference {steeg2014NIPS,
title = {Discovering Structure in High-Dimensional Data Through Correlation Explanation},
booktitle = {Advances in Neural Information Processing Systems, NIPS{\textquoteright}14},
year = {2014},
urlpaper = {http://arxiv.org/abs/1406.1222},
author = {Greg Ver Steeg and Aram Galstyan}
}

@conference {steeg2015corex_theory,
title = {Maximally Informative Hierarchical Representations of High-Dimensional Data},
booktitle = {AISTATS{\textquoteright}15},
year = {2015},
urlpaper = {http://arxiv.org/abs/1410.7404},
author = {Greg Ver Steeg and Aram Galstyan}
}

@book{Cover06,
author = {Cover, Thomas M. and Thomas, Joy A.},
title = {Elements of Information Theory (Wiley Series in Telecommunications and Signal Processing)},
year = {2006},
isbn = {0471241954},
publisher = {Wiley-Interscience},
address = {USA}
}

@article{Cichy21,
  author    = {Radoslaw Martin Cichy and
               Kshitij Dwivedi and
               Benjamin Lahner and
               Alex Lascelles and
               P. Iamshchinina and
               M. Graumann and
               Alex Andonian and
               N. A. R. Murty and
               K. Kay and
               Gemma Roig and
               Aude Oliva},
  title     = {The Algonauts Project 2021 Challenge: How the Human Brain Makes Sense
               of a World in Motion},
  journal   = {CoRR},
  volume    = {abs/2104.13714},
  year      = {2021},
}

@article{Field96,
  title={Emergence of simple-cell receptive field properties by learning a sparse code for natural images},
  author={Bruno A. Olshausen and David J. Field},
  journal={Nature},
  year={1996},
  volume={381},
  pages={607-609}
}

@article{DeAngelis97,
  title={Spatiotemporal receptive field organization in the lateral geniculate nucleus of cats and kittens.},
  author={D Cai and Gregory C. DeAngelis and Ralph D. Freeman},
  journal={Journal of neurophysiology},
  year={1997},
  volume={78 2},
  pages={1045-61}
}

@inproceedings{Watson92,
  title={DCT quantization matrices visually optimized for individual images},
  author={Andrew B. Watson},
  booktitle={Electronic Imaging},
  year={1993}
}

@article{Brainard19,
  title={The psychophysics toolbox},
  author={Brainard, David H and Vision, Spatial},
  journal={Spatial vision},
  volume={10},
  pages={433--436},
  year={1997},
  publisher={VSP}
}

@article{Esteve20,
  author = {Esteve, JJ. and Aguilar, G. and Maertens, M. and Wichmann, FA. and Malo, J.},
  title = {Psychophysical Estimation of Early and Late Noise},
  journal = {ar{X}iv 10.48550/arxiv.2012.06608},
  year = {2020},
}

@article{Malo2019FI, 
title={Information Flow in Biological Networks for Color Vision}, 
volume={24}, 
DOI={10.3390/e24101442}, 
number={10}, 
journal={Entropy}, 
publisher={MDPI AG}, 
author={Malo, Jesús}, 
year={2022}, 
month={Oct}, 
pages={1442} }

@INBOOK{Uriegas94,
  author =       {Martinez-Uriegas, E.},
  editor =       {Kelly, D H},
  title =        {Chromatic-achromatic multiplexing in human color vision},
  chapter =      {Chapt. 4 in Visual Science and Engineering: Models and Applications},
  pages =        {117-187},
  publisher =    {CRC Press},
  year =         {1994},
}

@article{Watson97,
  title={Model of visual contrast gain control and pattern masking},
  author={Watson, A. B. and Solomon, J. A.},
  journal={JOSA A},
  volume={14},
  number={9},
  pages={2379--2391},
  year={1997},
  publisher={Optical Society of America}
}

@article{Wainwright00,
author = {Rao, Rajesh and Olshausen, B and Lewicki, M and Wainwright, Martin and Schwartz, Odelia and Simoncelli, Eero},
year = {2001},
month = {01},
pages = {},
title = {Natural Image Statistics and Divisive Normalization: Modeling Nonlinearities and Adaptation in Cortical Neurons},
journal = {Statistical Theories of the Brain}
}

@ARTICLE{Schwartz01,
  AUTHOR = "O. Schwartz and E.P. Simoncelli",
  TITLE  = "Natural signal statistics and sensory gain control",
  JOURNAL = "Nature Neurosci.",
  VOLUME = 4,
  PAGES = "819--825",
  YEAR = 2001,
  NUMBER = "8"}

@ARTICLE{Malo01,
  author={Malo, J. and Gutierrez, J. and Epifanio, I. and Ferri, F.J. and Artigas, J.M.},
  journal={IEEE Trans. Im. Proc.}, 
  title={Perceptual feedback in multigrid motion estimation using an improved {DCT} quantization}, 
  year={2001},
  volume={10},
  number={10},
  pages={1411-1427}}

@article{Malo06a,
title={Nonlinear image representation for efficient perceptual coding},
author={Malo, J. and Simoncelli, E},
journal={IEEE Trans.Im.Proc.},
volume={15},
number={1},
pages={68--80},
year={2006},
publisher={IEEE}
}

@article{Malo06b,
  title={V1 non-linear properties emerge from local-to-global non-linear {ICA}},
  author={Malo, J. and Guti{\'e}rrez, J.},
  journal={Network: Computation in Neural Systems},
  volume={17},
  number={1},
  pages={85--102},
  year={2006},
  publisher={Taylor \& Francis}
}

@article{Malo10,
  title={Psychophysically tuned divisive normalization approximately factorizes the PDF of natural images},
  author={Malo, J. and Laparra, V.},
  journal={Neural computation},
  volume={22},
  number={12},
  pages={3179--3206},
  year={2010},
}

@article{Coen12,
    author = {Coen-Cagli, Ruben AND Dayan, Peter AND Schwartz, Odelia},
    journal = {PLOS Comp. Biol.},
    publisher = {Public Library of Science},
    title = {Cortical Surround Interactions and Perceptual Salience via Natural Scene Statistics},
    year = {2012},
    month = {03},
    volume = {8},
    pages = {1-18},
    number = {3},
}

@book{Kandel91,
  address = {New York},
  edition = {Third},
  editor = {Kandel, Eric R. and Schwartz, James H. and Jessell, Thomas M.},
  publisher = {Elsevier},
  title = {Principles of Neural Science},
  year = 1991
}

@article{Li19,
  title={A new framework for understanding vision from the perspective of the primary visual cortex},
  author={Li Zhaoping},
  journal={Current Opinion in Neurobiology},
  year={2019},
  volume={58},
  pages={1-10}
}

@INPROCEEDINGS{Watson02,
  author={A. B. {Watson} and J. {Malo}},
  booktitle={{IEEE} Proc. Int. Conf. Im. Proc.},
  title={Video quality measures based on the standard spatial observer},
  year={2002},
  volume={3},
  pages={III-III},
  }

@article{Malo97,
  title={Characterization of the human visual system threshold performance by a weighting function in the Gabor domain},
  author={Malo, J. and Pons, AM. and Felipe, A. and Artigas, JM.},
  journal={Journal of Modern Optics},
  volume={44},
  number={1},
  pages={127-148},
  year={1997},
  publisher={Taylor \& Francis},
}

@article{Watson94,
author = {Watson, Andrew},
year = {1994},
month = {08},
pages = {},
title = {Image Compression Using the {DCT}},
volume = {4},
journal = {Mathematica Journal},
}

@article{Ahumada97,
author = {Ahumada, Albert and Peterso, Heidi},
year = {1997},
month = {12},
title = {Luminance-Model-Based {DCT} Quantization for Color Image Compression},
volume = {1666},
journal = {Proc SPIE Human Vision, Visual Process Display III},
}

@article{Malo97subjective,
  title={Subjective image fidelity metric based on bit allocation of the human visual system in the {DCT} domain},
  author={Malo, J. and Pons, AM and Artigas, J.M. },
  journal={Im. Vis. Comp.},
  volume={15},
  number={7},
  pages={535--548},
  year={1997},
  publisher={Elsevier}
}

@article{camps-valls2008a,
  author  = {Camps, G. and Guti{{\'e}}rrez, J. and G{{\'o}}mez, G. and Malo, J.},
  title   = {On the Suitable Domain for SVM Training in Image Coding},
  journal = {J. Mach. Learn. Res.},
  year    = {2008},
  volume  = {9},
  number  = {3},
  pages   = {49--66},
}

@book{Fairchild13,
  title={Color Appearance Models},
  author={Fairchild, M.D.},
  series={The Wiley-IS\&T Series in Imag. Sci. Tech.},
  year={2013},
  publisher={Wiley}
}

@article{Laparra10,
  title={Divisive normalization image quality metric revisited},
  author={Laparra, V. and Mu\~noz, J. and Malo, J.},
  journal={JOSA A},
  volume={27},
  number={4},
  pages={852--864},
  year={2010},
  publisher={Optical Society of America}
}

@INPROCEEDINGS{Hepburn2020,
    author={A. {Hepburn} and V. {Laparra} and J. {Malo} and R. {McConville} and R. {Santos-Rodriguez}},
    booktitle={IEEE ICIP},
    title={Perceptnet: A Human Visual System Inspired Neural Network For Estimating Perceptual Distance},
    year={2020},
    volume={},
    number={},
    pages={121-125},
}

@article{Wang09,
    author = {Wang, Z. and Bovik, A. C.},
    journal = {IEEE Signal Processing Magazine},
    number = {1},
    pages = {98--117},
    title = {{Mean squared error: Love it or leave it? A new look at signal fidelity measures}},
    volume = {26},
    year = {2009}
}

@inproceedings{Teo94,
booktitle={IEEE ICIP},   
title={Perceptual image distortion},   
year={1994},
author={P.C. Teo and D.J. Heeger},
volume={2},  
pages={982-986},  
}

@book{Watson93,
editor = {Watson, Andrew B.},
title = {Digital Images and Human Vision},
year = {1993},
publisher = {MIT Press},
address = {Cambridge, MA, USA}
}

@article{Gutierrez06,
  title={Regularization operators for natural images based on nonlinear perception models},
  author={Guti{\'e}rrez, J. and Ferri, F. J and Malo, J.},
  journal={IEEE Trans. Im. Proc.},
  volume={15},
  number={1},
  pages={189--200},
  year={2006},
}

@book{Stark94,
author = {Stark, H. and Woods, J.},
year = {1994},
title = {Probability, Random Processes, and Estimation Theory for Engineers},
volume = {90},
journal = {Englewood Cliffs: Prentice Hall, 1986},
}

@article{Portilla01,
     AUTHOR= "J Portilla and E P Simoncelli",
     TITLE= "A parametric texture model based on joint statistics of
     complex wavelet coefficients",
     JOURNAL= "Int. J. Comp. Vis.",
     VOLUME= 40,
     NUMBER= 1,
     PAGES= "49--71",
     YEAR= 2000,
}

@article{Laparra20,
  author    = {Valero Laparra and
               Juan Emmanuel Johnson and
               Gustau Camps{-}Valls and
               Ra{\'{u}}l Santos{-}Rodr{\'{\i}}guez and
               Jesus Malo},
  title     = {Information Theory Measures via Multidimensional Gaussianization},
  journal   = {CoRR},
  volume    = {abs/2010.03807},
  year      = {2020},
  url       = {https://arxiv.org/abs/2010.03807},
  eprinttype = {arXiv},
  eprint    = {2010.03807}
}

@article{Semedo21,
author = {Semedo, João and Jasper, Anna and Zandvakili, Amin and Aschner, Amir and Machens, Christian and Kohn, Adam and Yu, Byron},
year = {2022},
journal={Nat Commun},
pages = {13},
volume={1099},
title = {Feedforward and feedback interactions between visual cortical areas use different population activity patterns},
}

@article{Kerkoerle14,
author = {Timo van Kerkoerle  and Matthew W. Self  and Bruno Dagnino  and Marie-Alice Gariel-Mathis  and Jasper Poort  and Chris van der Togt  and Pieter R. Roelfsema },
title = {Alpha and gamma oscillations characterize feedback and feedforward processing in monkey visual cortex},
journal = {PNAS},
volume = {111},
number = {40},
pages = {14332-14341},
year = {2014},
}

@article{Klink17,
title = {Distinct Feedforward and Feedback Effects of Microstimulation in Visual Cortex Reveal Neural Mechanisms of Texture Segregation},
journal = {Neuron},
volume = {95},
number = {1},
pages = {209-220.e3},
year = {2017},
author = {P. Christiaan Klink and Bruno Dagnino and Marie-Alice Gariel-Mathis and Pieter R. Roelfsema}
}

@article{Mejias16,
author = {Jorge F. Mejias  and John D. Murray  and Henry Kennedy  and Xiao-Jing Wang },
title = {Feedforward and feedback frequency-dependent interactions in a large-scale laminar network of the primate cortex},
journal = {Science Advances},
volume = {2},
number = {11},
pages = {e1601335},
year = {2016},
}

@article{Hulusi15,
author = {Kafaligonul, Hulusi and Breitmeyer, Bruno and Öğmen, Haluk},
year = {2015},
month = {03},
pages = {},
title = {Feedforward and feedback processes in vision},
volume = {6},
journal = {Front. Psychol.},
}

@article{szabo14,
author = {Z. Szab\'{o}},
title = {Information Theoretical Estimators Toolbox},
journal = {J. Mach. Learn. Res.},
year = {2014},
volume = {15},
pages = {283-287},
}

@article{Laparra12,
  title={Nonlinearities and adaptation of color vision from sequential principal curves analysis},
  author={Laparra, V. and Jim{\'e}nez, S. and Camps-Valls, G. and Malo, J.},
  journal={Neural Computation},
  volume={24},
  number={10},
  pages={2751--2788},
  year={2012},
  publisher={MIT Press 55 Hayward Street, Cambridge, MA 02142-1315 email: journals-info@ mit. edu}
}

@article{Gutmann14,
  title={Spatio-chromatic adaptation via higher-order canonical correlation analysis of natural images},
  author={Gutmann, M. U and Laparra, V. and Hyv{\"a}rinen, A. and Malo, J.},
  journal={PloS one},
  volume={9},
  number={2},
  pages={e86481},
  year={2014},
  publisher={Public Library of Science}
}

@article {Parraga09,
title = "Color Constancy Algorithms: Psychophysical Evaluation on a New Dataset",
journal = "Journal of Imaging Science and Technology",
year = "2009",
volume = "53",
number = "3",
pages = "31105-1-31105-9",
author = "Vazquez-Corral, J. and P{\'a}rraga, C. and Baldrich, R. and Vanrell, M.",
}

@article{Epifanio03,
  title={Linear transform for simultaneous diagonalization of covariance and perceptual metric matrix in image coding},
  author={Epifanio, I. and Gutierrez, J. and Malo, J.},
  journal={Patt. Recog.},
  volume={36},
  number={8},
  pages={1799--1811},
  year={2003},
  publisher={Pergamon}
}

@ARTICLE{Clarke81,
   author = {R.J. Clarke},
   title = {Relation between the Karhunen Loève and cosine transforms},
   journal = {IEE Proceedings F (Comm. Radar Sig. Proc.)},
   issue = {6},
   volume = {128},
   year = {1981},
   pages = {359-361},
}

@article{Barlow61,
author = {Barlow, H.},
year = {1961},
month = {01},
title = {Possible Principles Underlying the Transformations of Sensory Messages},
volume = {1},
journal = {Sensory Comm.},
}

@article{Messe06,
title = "Fixed or variable noise in contrast discrimination? The jury's still out...",
author = "Georgeson, {Mark A.} and Meese, {Timothy S.}",
year = "2006",
month = "11",
volume = "46",
pages = "4294--4303",
journal = "Vision Research",
publisher = "Elsevier",
number = "25",
}

\end{document}